\documentclass[twocolumn,secnumarabic,amssymb, nobibnotes, aps, prd,amsmath,amssymb,]{revtex4-1}

\newcommand{\elem}[2]{\ensuremath{{}^{#1}}#2}
\newcommand{\bra}[1]{\langle#1|}
\newcommand{\ket}[1]{|#1\rangle}

\usepackage{graphicx}
\usepackage{dcolumn}
\usepackage{bm}
\usepackage{multirow}
\usepackage{feynmf}
\usepackage{longtable}
\begin{document}

\title{
Finite-size effects and collective vibrations in the inner crust of 
neutron stars}%

\author{S.Baroni$^{a}$, A. Pastore$^{b}$, F.Raimondi$^{b}$, 
F.Barranco$^{c}$, R.A.Broglia$^{d,e,f}$
         and E.Vigezzi$^{e}$}%
\affiliation{
$^a$ TRIUMF, 4004 Wesbrook Mall, Vancouver BC, V6T 2A3, Canada\\
$^b$ Department of Physics, Post Office Box 35 (YFL), FI-40014 University of Jyv\"askyl\"a, Finland.\\
$^c$ Departamento de Fisica Aplicada III, Escuela Superior de Ingenieros, Camino de los Descubrimientos s/n,  
  41092 Sevilla, Spain.\\
$^d$ Dipartimento di Fisica, Universit\`a degli Studi di Milano, Via Celoria 16, 20133 Milano, Italy.\\
$^e$ INFN, Sezione di Milano, Via Celoria 16, 20133 Milano, Italy.\\
$^f$ The Niels Bohr Institute, University of Copenhagen, Blegdamsvej 17, 2100 Copenhagen \O, Denmark.
}
\date{\today}

\begin{abstract}
We study the linear response of the inner crust of neutron stars within the
Random Phase Approximation, employing a Skyrme-type interaction as effective interaction.
We adopt the Wigner-Seitz approximation, and consider a single unit cell of the 
Coulomb lattice which constitutes the inner crust, with a nucleus at its center,
surrounded by a sea of free neutrons.
With the use of an appropriate operator, it is possible to  
analyze in detail the properties of the vibrations of the surface of the nucleus 
and their interaction with the modes of the sea
of free neutrons, and to investigate the role of shell effects and of resonant states.
\end{abstract}
\maketitle

\section{Introduction}


In going from the outer crust of a neutron star towards the core of the star, one crosses the so-called inner crust. 
The baryonic density of the inner crust ranges from about $\rho = 10^{11}$ ${\rm g \; cm}^{-3}$ to approximately $\rho_{0}/2$, where $\rho_{0} = 2.8 \times 10^{14} \;{\rm g \; cm}^{-3}$ (or $ \rho_0 = 0.16$ fm$^{-3}$) is the density of nuclear matter at saturation.
We shall discuss the linear response of this system 
limiting ourselves to densities smaller than  $\rho_0/10$.  According to the generally accepted 
theoretical description the inner crust consists, in this region,  of a lattice of spherical nuclei 
immersed in a sea of free neutrons (with a background of electrons) \cite{Pet.Rav:95,Cha.Hae:08}. At these densities, neutrons are expected to be superfluid, and the inner crust of a rotating star should be threaded by vortex lines.

The presence of nuclei
affects a number of thermal and transport properties  of the inner crust,
and in some astrophysical scenarios may influence  the fast cooling of the star \cite{Pizzo,Monrozeau}. It can also affect the 
vortex dynamics, leading to vortex pinning, which might be the explanation
for the  phenomenon of glitches, sudden
changes observed in   the rotational period of pulsars \cite{vortexnoi1,vortexnoi2,Link}.

The interplay between the degrees of freedom of nuclei
and those of free neutrons has  been considered by different groups
 at the mean-field level within Hartree-Fock-Bogoliubov theory (HFB)
 \cite{Pizzo},\cite{Sandulescu,Sandulescu.ea:2004,Montani:2004}. 
It is well known, however, that many-body effects can
significantly modify the HFB results. Several studies 
have found that the pairing gap in neutron  matter is quenched by spin fluctuations
\cite{Cao,Schwenk,Gezerlis,Gandolfi}, but
results obtained in
uniform matter cannot be directly extrapolated to the actual case of the 
inhomogeneous inner crust. In fact, many-body effects in (isolated) atomic nuclei are strongly dominated by the coupling between single-particle and collective surface-like degrees of freedom. The exchange of surface modes between nucleons moving in time-reversal states 
close to the Fermi energy
gives rise to an ${\it attractive}$ interaction, which tends to
enhance  the pairing gap \cite{EPJ,Pastore}. Only exploratory calculations have 
directly addressed the induced interaction in the inner crust, finding that
the density fluctuations  associated with the dynamics of the nuclear surface
lead to a partial dequenching of the gap and to characteristic changes 
in the spatial dependence of the gap \cite{simone}.

Within this scenario 
it is of interest to investigate how the surface response 
changes from atomic nuclei (negative value of Fermi energy)
to the inner crust, where the neutron Fermi energy lies in the continuum.


In the present work 
we shall investigate the collective response of the nuclear surface 
by means of a  microscopic calculation of the linear response of the system based on the Random Phase
Approximation (RPA) theory \cite{Ringschuck}.  
 As in most previous studies, we shall deal with this inhomogeneous system within the Wigner-Seitz approximation (WS),
enclosing the system in a spherical unit cell. In this way, we shall neglect
the band structure associated with the Coulomb lattice \cite{Chamel}.
 
Previous studies \cite{Gori,Khan,Grasso} have shown that
the response to operators of the type  $r^{L} Y_{LM}$ or $[r^L Y_{L} \times \sigma]_{JM}$ 
is very close to that found in neutron matter, in keeping with the fact that
the nucleus occupies only a small fraction of the  volume of the WS cell.
However, in the following we shall focus on  the modes which are specifically 
associated with  the presence of the nucleus at the center of the WS cell. 
Therefore,  we shall study the response to the operator 
$dU/dr Y_{LM}$ , where $U$ is the mean-field potential.
In fact, $dU/dr Y_{LM}$ is the genuine operator associated with collective surface modes,
which appears naturally in the particle-vibration coupling Hamiltonian \cite{Bohrmott}.
Essentially the same results would be obtained using the density instead of the mean-field
potential. In Appendix B we also briefly consider the operator $r^{L} Y_{LM}$.

In this work  we shall not consider the effects of superfluidity on the linear response. 
Pairing correlations can affect considerably the low-lying part of the spectrum, 
not only because they smear the Fermi surface  leading to the partial occupation of single-particle states, but also because they can change the isotopic composition of the crust \cite{Baldo}.  We shall focus on the properties of giant resonances, which are not expected 
to be sensitive to pairing. Within this context, we shall use in the calculations
the favoured number of protons
predicted by  Negele and Vautherin \cite{Neg.Vau:73}, predictions 
which were worked out  without pairing. 
We have anyway indicated  in the text where one expects to see the main  effects of pairing
correlations.


\section{Outline  of the  calculations}

We start from a mean-field Hartree-Fock (HF)  calculation,
making use of the Skyrme-type SLy4 interaction \cite{Cha.ea:98a} and working
on a grid with a mesh size of 0.1 fm.
Our calculation is in line with  other studies 
previously performed by a number of  groups \cite{Pizzo,Sandulescu.ea:2004,Montani:2004}.
We impose that the single-particle wavefunctions
vanish at the edges of the WS cell. 
We expect that within the selected density 
range the results should not be sensitive
to the particular choice of the boundary conditions
(cf. also Appendix A). 
For densities larger than those considered in this paper,
the  WS approximation starts to break down, leading to a dependence on the 
boundary conditions \cite{Baldo}.

We shall show results for two WS cells. 
The first one has a radius 
$R_{WS}$= 33.1 fm
and contains 1314 neutrons and 50 protons ($^{1364}$Sn), corresponding to nucleon density 
$\rho \approx 1.5 \times 10^{13}$ g cm$^{-3} $ (or $0.05 \rho_0$).
Far from the nucleus the density approaches the asymptotic density 
$\rho_{\infty} \approx 0.01$ fm  $^{-3}$.
The second one  has a radius $R_{WS} = $ 42.2 fm and contains 458 neutrons
and 40 protons ($^{498}$Zr), corresponding to the nucleon
density $\rho \approx 2.7 \times 10^{12}$ g cm$^{-3} $ (or $0.01 \rho_0$)
and to $\rho_{\infty} \approx 1.5 \times 10^{-3}$ fm$^{-3}$.
The number of protons and neutrons and the cell radius are an input to the HF calculation and  have been taken from  the results of Negele and Vautherin \cite{Neg.Vau:73}. 
They found that the favoured proton number is equal to $Z=40$ in the lower-density
part of the inner crust, changing to $Z=50$ for densities larger than  
$\rho \approx 3 \times 10^{12}$ g cm$^{-3}$. Other authors have determined the isotopic 
composition of the  crust using different energy functionals and different methods.
In particular, systematic calculations 
have been performed in ref. \cite{Baldo} including pairing correlations,
neglected by Negele and Vautherin.   
The results of these studies  have been summarized in ref. \cite{Cha.Hae:08}. In most cases, one finds that  the favoured proton number varies between $Z \approx $ 35 and $ Z = 50$, becoming smaller ($Z \approx 25 $) in the deeper regions of the crust.   These uncertainties are
understandable, because typical energy differences between local minima in the energy
as a function of the proton number at a given density are of the order of 10 keV/nucleon. 
In the following, we shall also consider the WS cell $^{506}$Sn, calculated with the
same parameters as $^{498}$Zr, but for the different proton number  $Z=50$
instead of $Z=40$.

The corresponding Fermi energy
for neutrons and protons are equal to $E_F = 5.5 $ MeV and $E_F =-35.1$ MeV 
for $^{1364}$Sn,  and to $E_F= 2.1 $ MeV and $E_F = -31.5 $ MeV  for $^{498}$Zr.
We note that the potential reaches an almost constant value far from the nucleus, so that the
continuum portion of the single-particle spectrum starts at about
$E_{cont}= -3.5 $ MeV ($^{1364}$Sn) and at $E_{cont}= -0.5$ MeV ($^{498}$Zr).
The number of neutrons in bound orbitals is about the same in
the $^{1364}$Sn (126 bound neutrons) and in the $^{498}$Zr (124 bound neutrons) cases.

\begin{figure*}[!ht]
\begin{center}
  \includegraphics[width=0.36\textwidth,angle=-90]{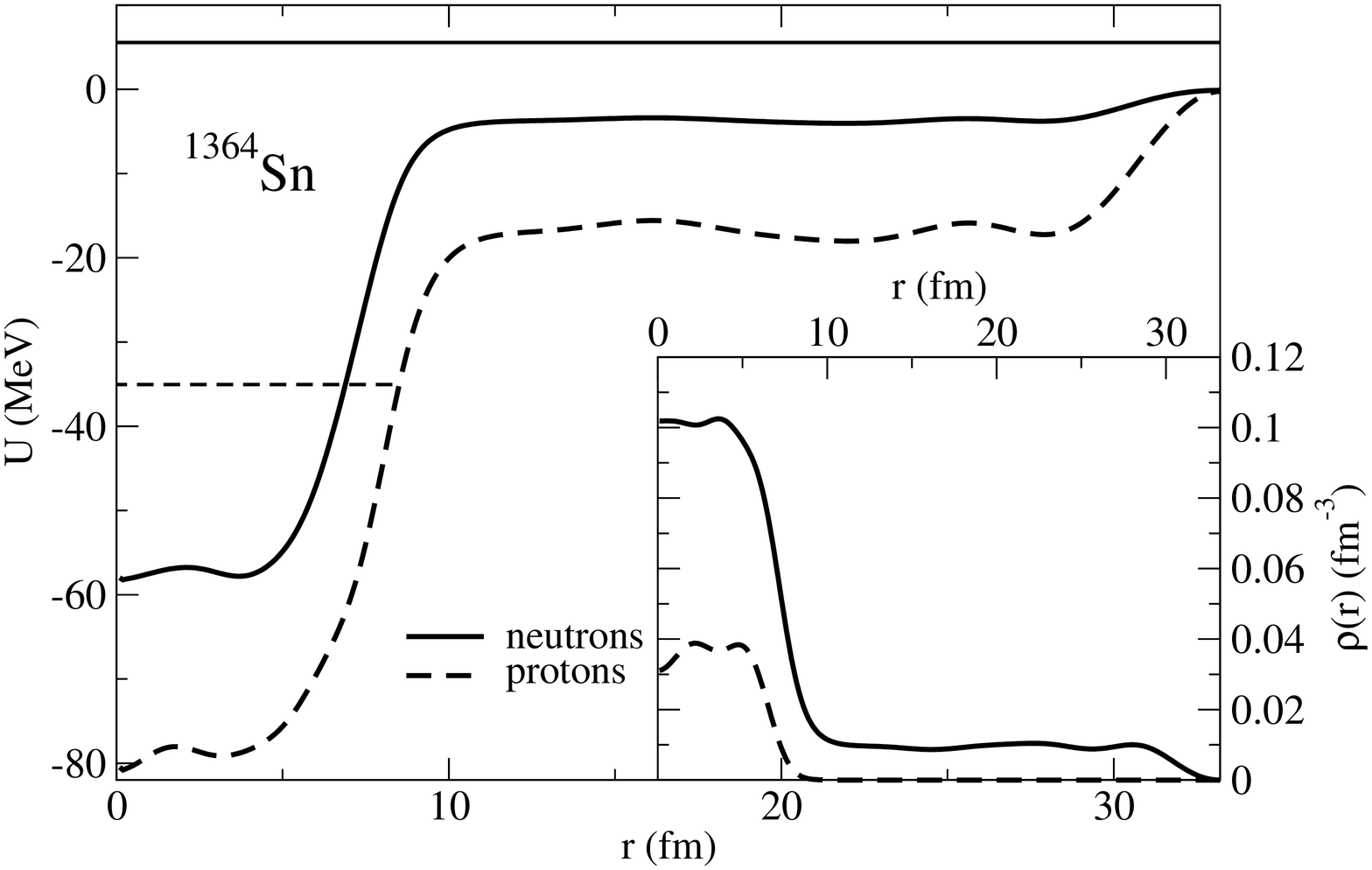}
  \includegraphics[width=0.36\textwidth,angle=-90]{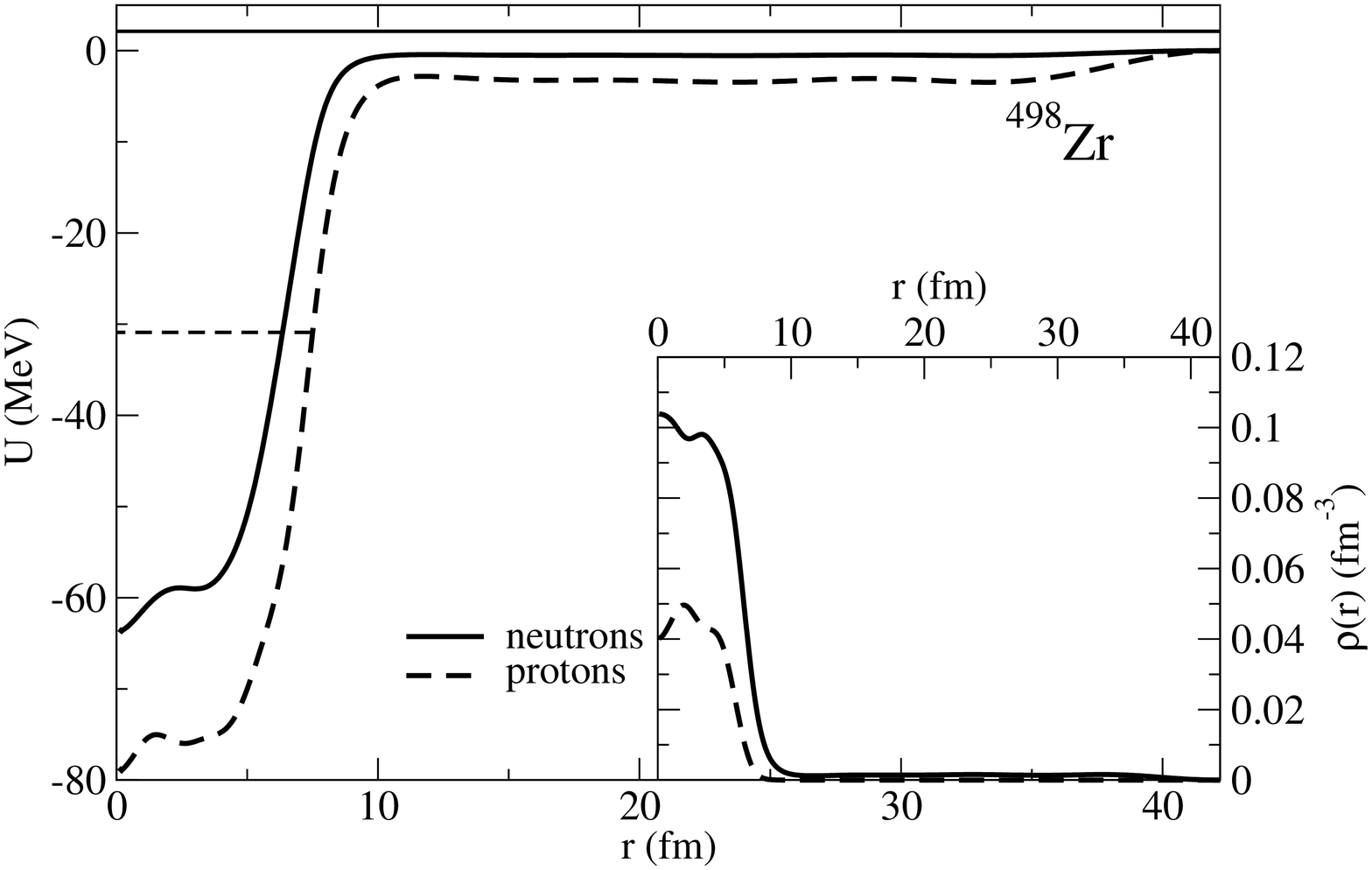}
\end{center}
 \caption{(left) Mean-field potential  calculated in the  WS cell \elem{1364}{Sn}.
(right) The same, for the WS cell \elem{498}{Zr}. The thin horizontal lines 
indicate the neutron and the proton Fermi energies, while in the insets we show the local densities.
}
\label{IND HFRPA:fig 10}
\end{figure*}

In Fig.\ref{IND HFRPA:fig 10} we show the mean-field potential $U(r)$
in the two cases \cite{Brink}. The presence of the outer neutron gas leads to a more diffuse
potential than that found in atomic nuclei.  The neutron potentials
can be parametrized as Woods-Saxon potentials,
\begin{equation}
U(r) = U_{\infty} + \frac {U_0}{1 
+ {\rm exp} \frac{r - R_0}{a}},
\end{equation}
with the values 
$U_{\infty} = $ -3.73 MeV, $U_0$= -54.2 MeV, 
$R_0 = 7.06$  fm, $a = 0.79$ fm 
for $^{1364}$Sn and
$U_{\infty} = $ -0.42 MeV,
$U_0$= -61.4 MeV,  
$R_0 = $ 6.42 fm, $a = 0.73$ fm 
for $^{498}$Zr.  
We remark that the values of the radii are close to 
those of ordinary nuclei with the same number
of bound nucleons  (1.2 $A^{1/3} \approx 7$ fm for
$A \approx 200$ ). The extension of  
proton and neutron potentials is similar.
The radius of $^{498}$Zr is 
close to the value calculated for  the drip-line nucleus 
$^{176}$Sn (cf. below Section III(c)). We can use the 
value of the radius $R_0$ to estimate the typical 
oscillator frequencies associated with the mean-field
potentials  shown in Fig.\ref{IND HFRPA:fig 10}. 
Using the expression $\hbar \omega \approx 75 N^{1/3}/R_0^{2}$
(which reduces  to the usual estimate $\hbar \omega \approx 40 A^{-1/3}$
for $N = Z$ and $R_0 = 1.2 A^{1/3}$), one finds
$\hbar \omega_n = $ 7.7 MeV for neutrons (using the number of 
bound neutrons, $N \approx 150$)
and   $\hbar \omega_p = $ 5.3 MeV 
for protons in $^{1364}$Sn. Taking into account that the effective mass $m_k$ 
associated with the HF potential increases the level spacing, one can
estimate $\hbar \omega_n = $ 9.6 MeV and $\hbar \omega_p = $ 6.6 MeV,
using $m_k/m=$ 0.8 (the typical 
value of the effective mass associated with the SLy4 interaction
in the nuclear surface region). We note  the small value of 
$\hbar \omega_p$, associated with the low proton density.

We have then calculated the excited states of the system performing
a RPA calculation in configuration space, based on the HF mean-field.
The residual interaction  was
taken as the second derivative of the energy functional,
that is, it was calculated with the same two-body Skyrme force which determines
the mean-field. The only terms that were dropped  in the p-h
channel were the two-body Coulomb and two-body spin-orbit. Their importance was
recently discussed in  refs. \cite{Tap.ea:2006,Peru} for atomic nuclei.
While they play a role in a detailed analysis of the 
experimental data, they hardly change the 
general properties of nuclear response.

The resulting  RPA phonons are classified according to their total angular momentum $J$ and parity
$\pi$, while the index $\nu=1,2,\dots$ runs over the states having the same spin and parity with increasing
energy. We have tested  the numerical scheme considering the case of WS cells without protons.
In this case, the nuclear potential is flat except at the edge of the cell, and the WS cell represents a portion
of neutron matter. The calculated response can then be compared with the analytical results obtained
in uniform matter. The corresponding results are collected in Appendix A.

\vspace{1cm} 

\begin{figure*}[!ht]
\centering
\includegraphics[width=0.36\textwidth]{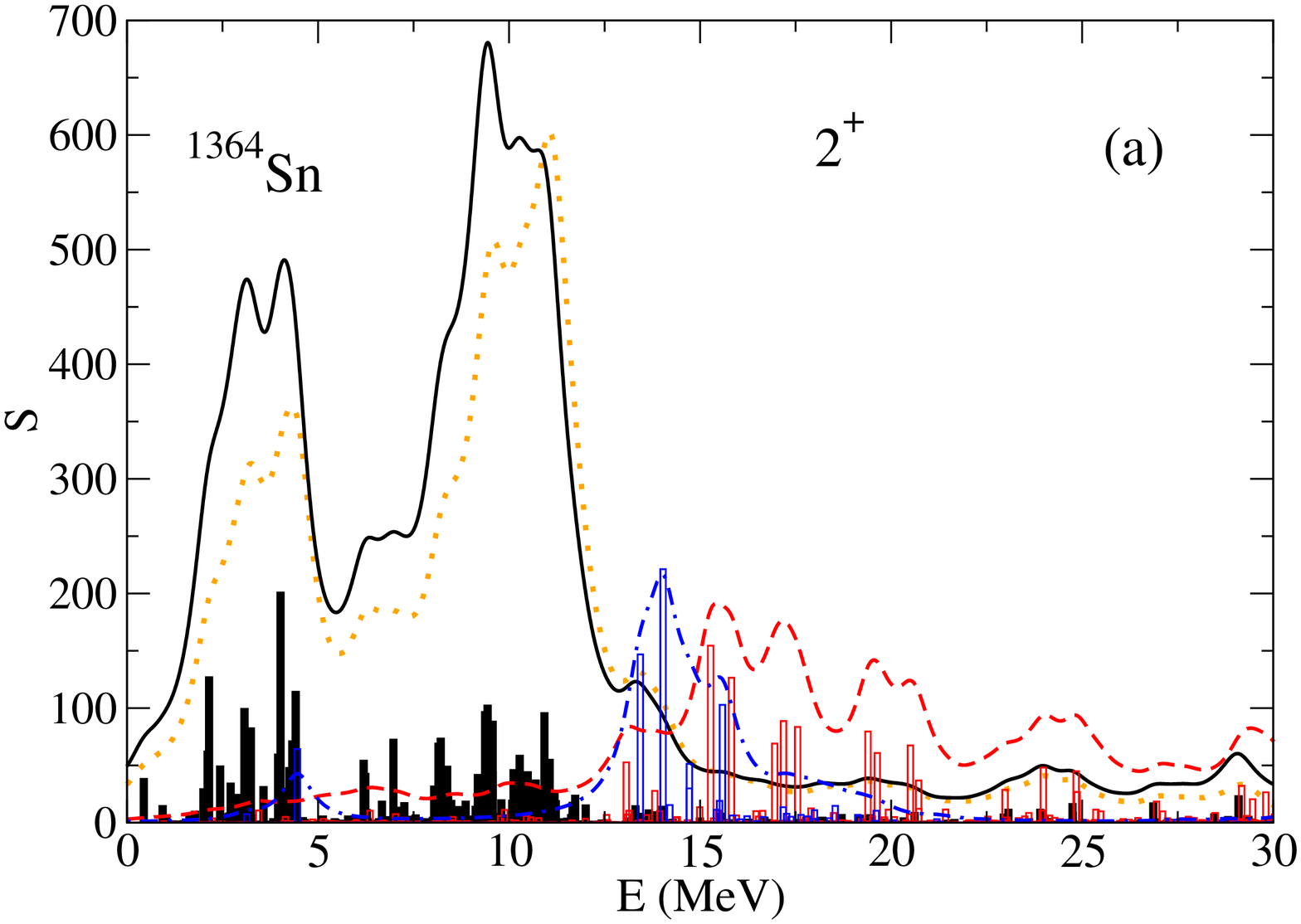}
\includegraphics[width=0.36\textwidth]{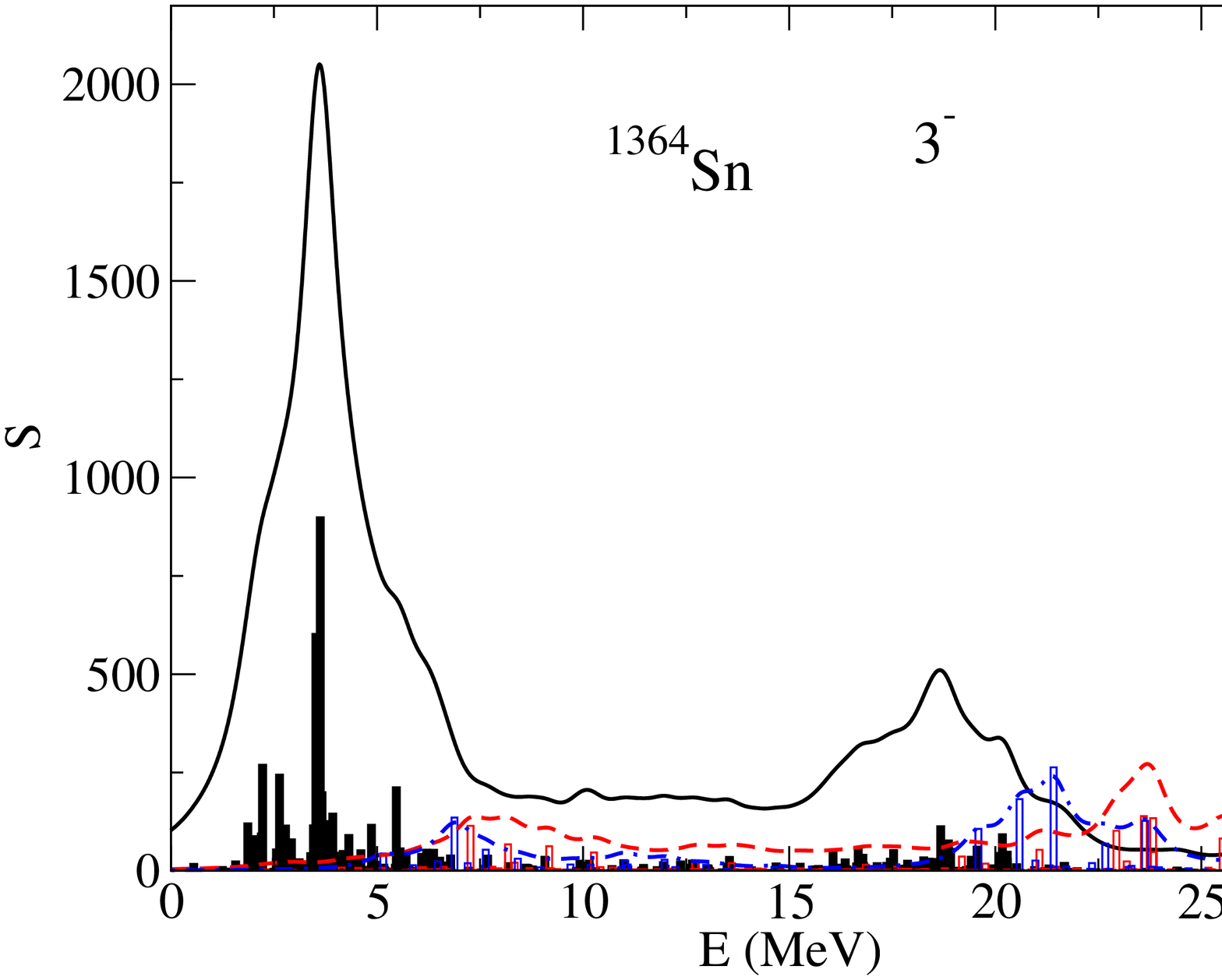}
\caption{ Quadrupole (a) and octupole (b) strength functions calculated in the $^{1364}$Sn WS cell.
Black histograms refer to the discrete RPA response, while red and blue histograms refer to the discrete HF response for neutron and protons respectively (in units of MeV$^2$ fm$^{-2}$).
The solid curve  (in units of MeV fm$^{-2}$) 
is  obtained  by a convolution of the discrete RPA strength 
with a Lorentzian curve having a FWHM
equal to 1 MeV. The dashed and dash-dotted curve are obtained convoluting 
the neutron and proton discrete HF strength respectively.
The dotted line in (a) shows the quadrupole RPA response obtained with $E_{cut}$ = 30 MeV (see text for details).}
\label{fig:1364_response}
\end{figure*}

\section{Results}

As discussed in the introduction, in the present  work we are particularly interested
in the study of the dynamics of the nuclear surface. 
For this reason
we shall 
study the response to the operator $dU(r)/dr Y_{LM}$, for $L$=2 and 3, where $U(r)$ represents  the HF potential 
calculated in the WS cell (cf. Fig. \ref{IND HFRPA:fig 10}).
This operator is specifically sensitive to the nuclear surface.
In the case of an isolated nucleus it leads to 
results similar to those obtained with the 
 $r^LY_{LM}$ operator for both quadrupole 
and octupole modes.
In the following we shall give a detailed discussion of the results obtained for 
the $^{1364}$Sn WS cell. We shall then present  the results associated with 
the $^{498}$Zr case. Finally, we shall consider the evolution of the response
in the  Sn isotopes with increasing number of neutrons, going  
from the closed-shell nucleus  $^{132}$Sn to the drip-line nucleus $^{176}$Sn and then 
into the inner crust to the WS cells $^{506}$Sn and $^{1364}$Sn.



\subsubsection*{(a) The case of $^{1364}$Sn}

In Fig. \ref{fig:1364_response} we show the HF and RPA strength functions
of the operator $dU/dr Y_{LM}$  in the $2^+$ and $3^-$
channels for the $^{1364}$Sn WS cell. The bars indicate the solution 
of the discrete RPA equations, while the continuous lines are obtained
folding the discrete response with a Lorentzian function having a 
Full Width at Half Maximum (FWHM) equal to 1 MeV. 

\begin{figure}
\begin{center}
\includegraphics[width=0.36\textwidth]{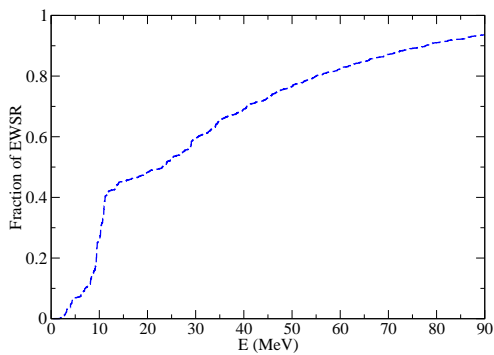}
\caption{Fraction of the quadrupole EWSR obtained integrating  the RPA strength 
function shown in Fig. \ref{fig:1364_response}(a) up to a given energy.}
\label{ewsr:1364sn}
\end{center}
\end{figure}

In order to  satisfy the Energy Weighted Sum Rule (EWSR) 
within 93\%  we had  to include single-particle states up to $E_{cut}=$ 90 MeV.
The $|ph\rangle$ basis is composed by about 8000 states for the quadrupole response  
and by about  10000 for the octupole response; these values  
are close to the limits of our computational possibilities. 
The fraction of the EWSR obtained by integrating  the calculated quadrupole response 
up to a given energy is
shown in Fig. \ref{ewsr:1364sn} as a function of energy.
We notice that  the main features of the low-lying part 
of the response can be  obtained with a much lower value of $E_{cut}$. This can be seen
in Fig.   \ref{fig:1364_response}(a), where the dotted line shows
the RPA response  obtained with $E_{cut}$= 30 MeV.




The proton unperturbed  response involves deeply bound orbitals. The
strongest particle-hole transitions connect orbitals separated by two
oscillator shells, and have an energy of approximately  14 MeV,
close to the value of $2 \hbar \omega_p$ estimated above.
There is also a rather strong  low-lying transition, $g_{9/2} \to d_{5/2}$,
whose energy is about 5 MeV. 

The energies and transition strengths associated with the strongest 
(unperturbed) neutron
particle-hole transitions 
are collected in Tables I (quadrupole response)  and II (octupole response).
They involve either bound states ($E_{lj} < E_{cont}= -5.5 $ MeV)
or states in the continuum ($E_{lj} >= E_{cont}$).
The latter are found to be resonant states, that is quasibound levels 
whose wavefunction is concentrated in the interior of the nucleus.
To show this we take a potential which is equal to the  mean-field potential
up to 30 fm and is constant and equal to $E_{cont}$ from 30 fm to infinity.
We then compute the resonant states associated to this potential, looking at
the wavefunction phase shift that occurs when a particle at a given energy
goes past the potential well.
The quantum numbers $\{l,j\}$ label a resonance,
if the associated phase shift
$\delta_{lj}(E)$ reaches the value
$(n+1) \pi/2$  $(n= 0,1,..)$ with a positive slope
at the energy $E_{res}$. 
The width of the resonance is taken to be equal
to the full width at half maximum (FWHM) of the derivative of $\delta_{lj}$\cite{bianchini}.
For narrow resonances, the resulting value is almost the same as that
obtained making use of  the expression
\begin{equation}
\Gamma_{res} = \frac{2}{\left ( d \delta_{lj}/dE\right )_{res}}.
\label{reso}
\end{equation}




Comparing Table \ref{table:dudr_1364sn_trans_2} and Table \ref{table:dudr_1364sn_trans_3}
with Table   \ref{table:dudr_1364sn_reso}, one realizes that the quantum numbers of the 
particle states coincide with one of those listed in Table   \ref{table:dudr_1364sn_reso}
(with the exception of the state $(l,j)= (9,17/2)$ in Table \ref{table:dudr_1364sn_trans_3}).
The wavefunctions of the resonant states in the interior of the nucleus
are similar to those of bound states belonging to the oscillator shells $N = 6-10$. 
Their angular momentum is the largest available in each shell, except for the 
shell $N$ = 6, where besides the state $l=6, j=11/2$ one  also finds the states 
$l=4, j=7/2$ and $l=4, j=9/2$. For the highest shells  ($N = 9,10$) only the state with $l = j+1/2$ 
has a resonant character, while its
spin-orbit partner $l = j-1/2$ lies at a too high energy respect to the 
centrifugal barrier.
The separation between resonances 
can be estimated from  Table \ref{table:dudr_1364sn_reso} to be 
$\hbar \omega_n \approx 8-9$ MeV, close to the value estimated above 
based on the shape and the effective mass of the HF potential.

The value of the   spin-orbit splitting for these large 
values of $l$ is close to $\hbar \omega_n$. 
As a consequence, the intruder state $(l,j+1/2)$ lies rather close to the state $(l-1, j-1/2)$, 
and the neutron part of the unperturbed response is characterized by strong 
peaks corresponding to  transition energies of the order of $2 \hbar \omega_n$ (quadrupole response) 
or of the order of  $1 \hbar \omega_n$ and $3 \hbar \omega_n$ (octupole response),
with a spreading caused by the width of the resonances.  In other
words, the strength function is determined  to a large extent by the shell structure, as in atomic nuclei.
This point will be further discussed below (cf. Figs. \ref{fig:dudr_evolution}
and \ref{fig:dudr_shell_evolution}).

\begin{figure*}[!h]
\begin{center}
\includegraphics[width=0.36\textwidth]{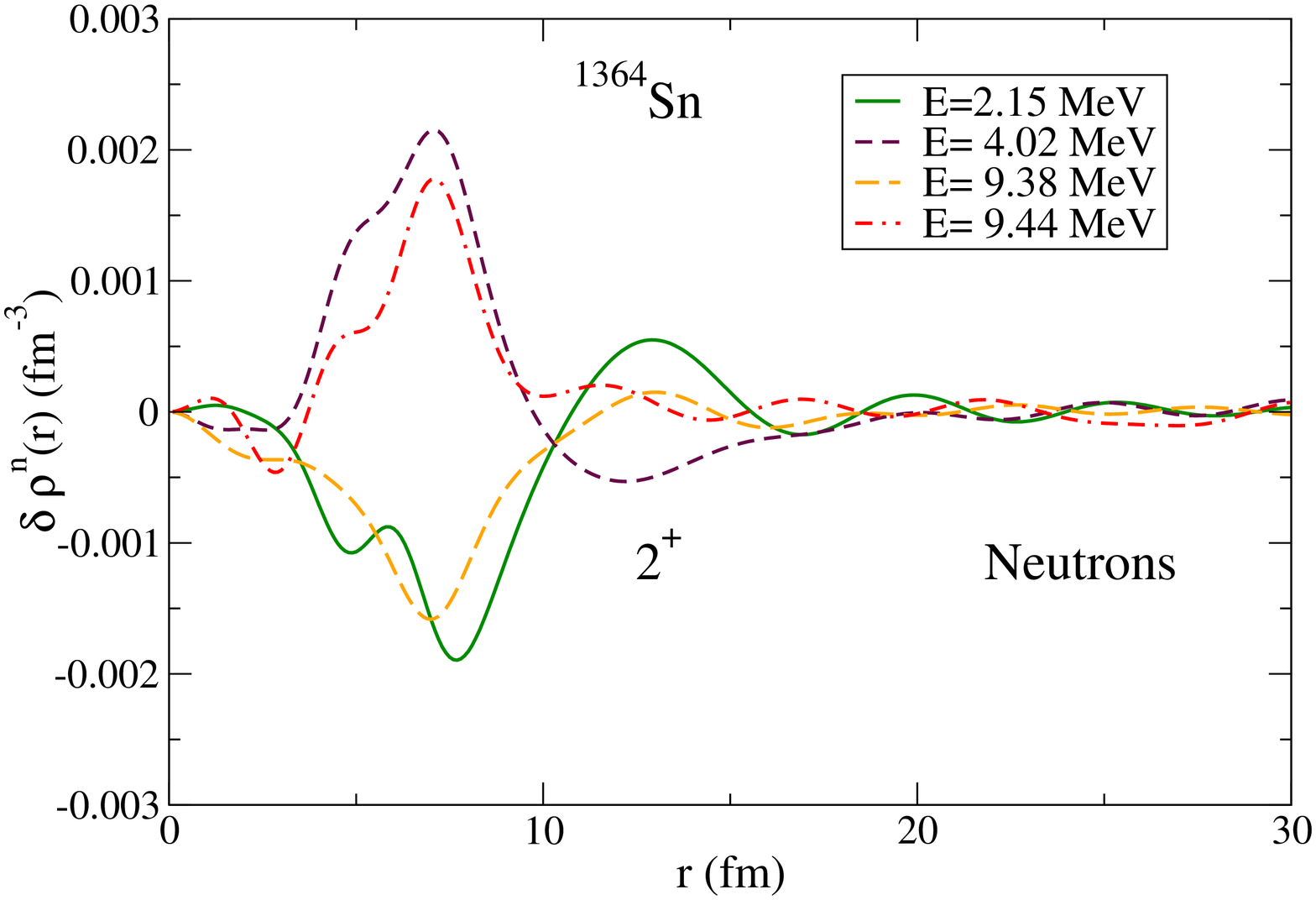}
\includegraphics[width=0.36\textwidth]{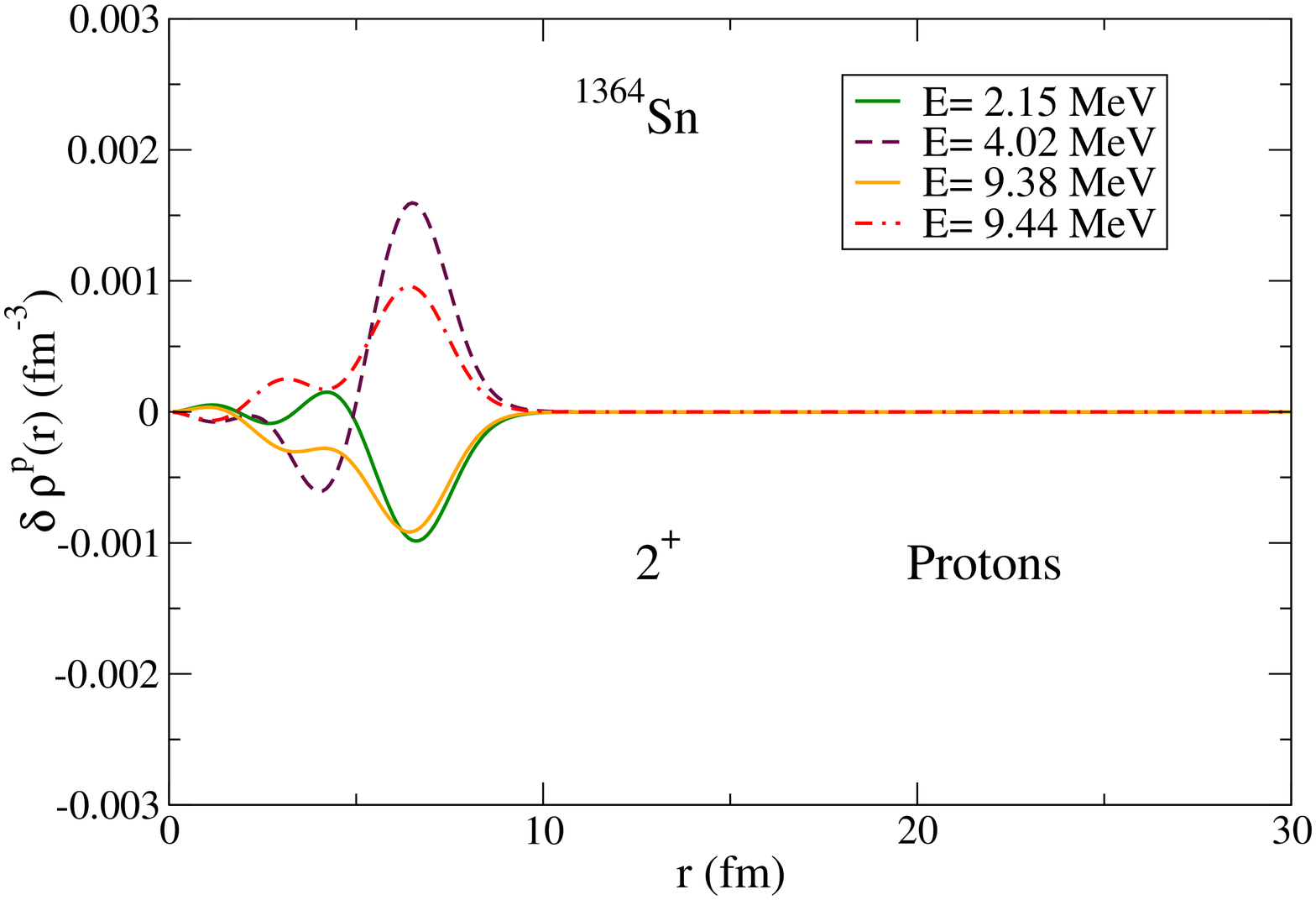}
\end{center}
\caption{Neutron (left panel) and proton (right) transition densities associated with the four strongest 
RPA transitions calculated in the response  
to the operator $dU/dr Y_{2M}$ for the $^{1364}$Sn  WS cell, shown in 
Fig. \ref{fig:1364_response}(a). 
The energies of the transitions are listed in the legends of the figures.
 }
\label{fig:dudr_1364sn_trans_2}
\end{figure*}

The RPA $2^+$ response, shown in Fig. \ref{fig:1364_response}(a), 
displays two main peaks,
lying close to 3.5 MeV and to 10.0 MeV, with a width of
about 3.3 and 4.0 MeV respectively.
The $3^-$  response  (cf.  Fig. \ref{fig:1364_response}(b)) shows a strong peak 
around 3.6 MeV with a width of about 2 MeV and a broader  high-energy bump
around 18.6 MeV with a width of about 4 MeV.
The neutron and proton transition densities associated with the four RPA
solutions carrying the largest quadrupole strength are shown 
in Fig. \ref{fig:dudr_1364sn_trans_2}. They are peaked at the surface of the nucleus,
and their shape is close to the derivative of the density,  
as   predicted by the  semiclassical model for surface collective modes.

\begin{figure*}[!h]
\begin{center}
\includegraphics[width=0.36\textwidth]{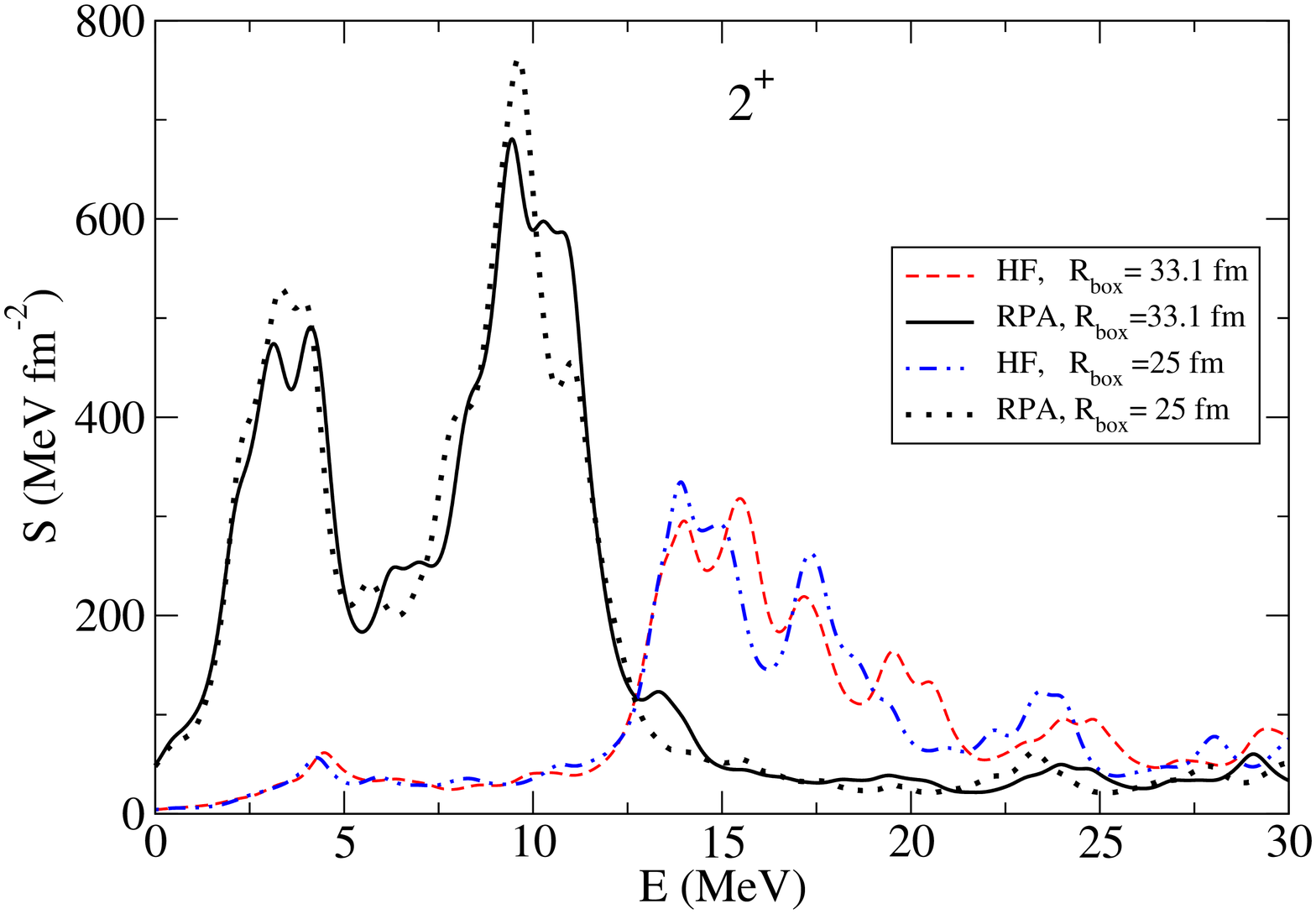}
\end{center}
\caption{ The HF (dashed line) and RPA (solid line) quadrupole strength functions calculated in $^{1364}$Sn,
are compared to those calculated in  a  WS cell
with 50 protons and  654 neutrons and a radius $R_{box} = $25 fm, corresponding to
about the same asymptotic neutron density (dashed and dash-dotted lines).}
\label{fig:dudr_1364_reducedbox}
\end{figure*}

The key role played by resonant transitions can be demonstrated  by performing a RPA calculation
in a WS cell with the same number of protons and with approximately the same asymptotic neutron  density,
 but with a reduced value of $R_{box} = 25$ fm.
The resulting strength, shown in Fig. \ref{fig:dudr_1364_reducedbox}  is very similar
to that obtained in $^{1364}$Sn.

\begin{figure*}[!h]
\begin{center}
\includegraphics[width=0.36\textwidth]{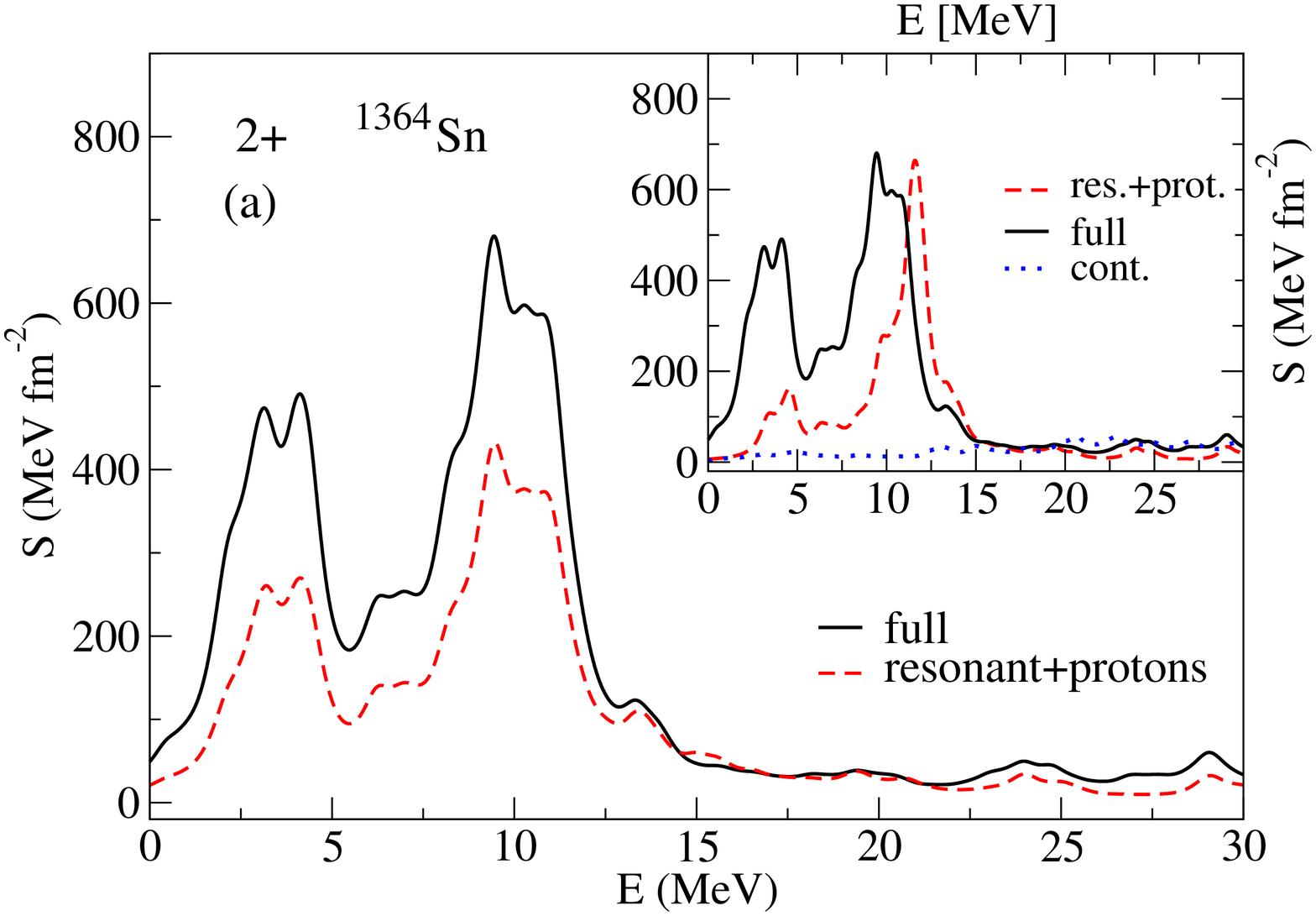}
\includegraphics[width=0.36\textwidth]{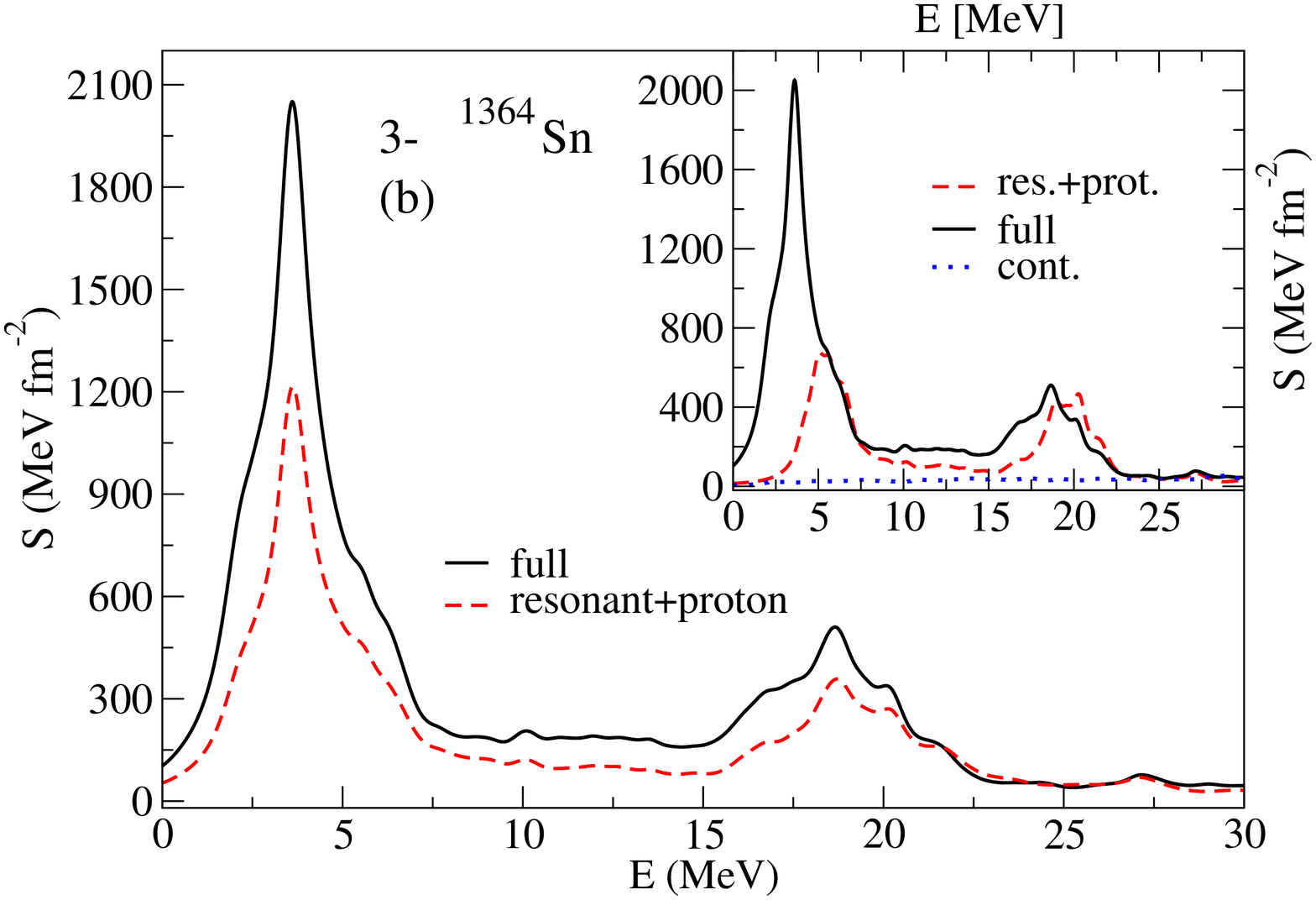}
\end{center}
\caption{(a) The RPA quadrupole strength function calculated for the $^{1364}$Sn  WS cell
(solid line, cf. Fig.\ref{fig:1364_response}(a)), is compared with the strength 
obtained including only the amplitudes associated with proton and  resonant neutron 
particle-hole transitions (dashed line, cf. text for details).
In the inset, the full RPA  quadrupole strength function is instead
compared with the RPA strength
calculated using a restricted particle-hole basis limited either 
to proton states and to bound and resonant neutron states (dashed line),
or to  neutron continuum states (dotted line)
(b). The same as (a), for the operator $dU/dr Y_{3M}$.}
\label{fig:dudr_1364sn_restricted_23}
\end{figure*}

In order to study in more detail  the resonant neutron 
particle-hole transitions in the RPA response, 
it is useful to distinguish their contributions to  the  strength from those 
involving the other, 'continuum' states \cite{Liotta}. We shall limit ourselves to a 
simple analysis and consider a single-particle
state in the WS cell to be 'resonant' if its quantum numbers $(l,j)$ coincide with one of those
listed in Table \ref{table:dudr_1364sn_reso}, 
and if its energy lies in the range 
$E_{lj}^{res} \pm \Gamma_{lj}^{res}$. 
The strength of multipolarity $L$
associated with a given phonon $|\nu>$ can be schematically written

\begin{eqnarray} 
S^{\nu}_L = |\sum_{ph} (X^{\nu}_{ph} + (-1)^L Y^{\nu}_{ph}) & \nonumber\\
<jp||Y_L||jh> <R_p|dU/dr|R_h>|^2&,
\label{sum}
\end{eqnarray}
where $X^{\nu}_{ph}$ and $Y^{\nu}_{ph}$ denote the forwardsgoing
 and backwardsgoing amplitudes 
associated with  each RPA root, while $R_p$,$R_h$ denote the single-particle radial
wavefunctions. 
In the following, we shall single out the 'resonant' contributions to the
computed strength function, restricting the sum in Eq. (\ref{sum}) 
to transitions between two bound states (including both protons and neutrons), 
and to 'resonant' neutron transitions,
namely those for which  at least 
one of the two states has a 'resonant' character.
In Fig. \ref{fig:dudr_1364sn_restricted_23}  we compare 
the full $2^+$ and $3^-$  strengths with the correspondent resonant contributions.
It is seen that the latter reproduce quite well the shape of the 
full response.


It is also interesting to 
perform a new RPA calculation excluding 
the transitions between bound states and  the resonant transitions
from the diagonalization.
This produces 
a featureless response which accounts for about $5\%$ 
of the EWSR 
$2^+$ in the range 0-20 MeV (cf. the dotted lines in the insets of Fig. \ref{fig:dudr_1364sn_restricted_23}). 
Performing instead a new RPA calculation including in the diagonalization
only the transitions between bound states and  the resonant transitions 
one finds that  the resulting 
quadrupole strength displays a very collective peak 
at $E \approx 12 $ MeV with a narrow width of the order of 2 MeV,
which accounts for about 30$\%$ of the EWSR
(cf. the dashed line in the inset of Fig. \ref{fig:dudr_1364sn_restricted_23}(a)). 
A small peak around 4.5 MeV is also visible. This originates from  low-energy transitions between 
the well-bound proton states. The octupole strength instead displays two peaks of comparable strength
located at about 6 MeV and 20 MeV. 
The coupling with continuum states 
increases the strength of the peaks and lowers their energy, producing the 
full strength:
in macroscopic terms, this can be associated with a decrease of the surface tension,
taking place when the nucleus is immersed in the neutron fluid, as well as
with the increase of the mass of the fluid which  takes part in the motion
\cite{liquiddrops}. 

\begin{table}
\begin{center}
\begin{tabular}{cccc|cccc|cc}
\hline
$n_h$& $l_h$ & $j_h$ & $E_h$ & $n_p$ & $l_p$& $j_p$ & $E_p$ & $E_{ph}$&  $T$[MeV$^2$fm$^{-2}$]\\ 
\hline
1& 5& 9/2       & -9.3 & 5&  7&  13/2 &    8.2 & 17.5  & 83.5\\ 
1& 5& 9/2       & -9.3 & 6&  7&  13/2 &   10.3 & 19.6  & 60.8\\
1& 6& 13/2      & -5.8 & 5&  8&  17/2 &    9.5  &15.3  &154.4\\
1& 6& 13/2      & -5.8 & 6&  8&  17/2 &   11.4  &17.2  & 88.7\\
2& 6& 11/2      &  0.1 & 7&  8&  15/2 &   17.0  &16.9  & 69.0\\
2& 6& 11/2      &  0.1 & 8&  8&  15/2 &   20.6  &20.5  & 67.4\\
3& 7& 15/2      &  2.2 & 7&  9&  19/2 &   18.1  &15.9  &126.4\\
3& 7& 15/2      &  2.2 & 8&  9&  19/2 &   21.6  &19.4  &79.4\\
\hline 
\end{tabular} 
\caption{List of the eight unperturbed neutron particle-hole transitions 
in the $^{1364}$WS cell and 
associated with the largest transition strengths calculated 
with the operator $dU/dr Y_{2M}$. 
In the first four columns we give principal quantum number, the orbital angular momentum, the total angular momentum and the energies  $n_h, l_{h},j_h$ and $E_h$ of the hole; in the 
next four columns, the  corresponding quantities $n_p,l_{p},j_p$ and $E_p$
for the particle. In the last two columns, we give the energy 
$E_{ph} = E_p - E_h$ and the transition strength 
$T$ associated with the transition. All energies are in MeV.}
\label{table:dudr_1364sn_trans_2}
\end{center}
\end{table}

\begin{table}
\begin{center}
\begin{tabular}{cccc|cccc|cc}
\hline
$n_h$& $l_h$ & $j_h$ & $E_h$& $n_p$ & $l_p$& $j_p$ & $E_p$& $E_{ph}$&  $T$[MeV$^2$fm$^{-2}$]\\ 
\hline
1& 5& 11/2      & -14.1 & 5&  8&  17/2 &    9.5 &23.6& 139.0\\
1& 5& 11/2      & -14.1 & 6&  8&  17/2 &   11.4 &25.5&  82.2\\
1& 5&  9/2      & -9.3 &  8& 8&  15/2 &   20.6 &29.9&  67.4\\
1& 6& 13/2      & -5.8 &  7& 9&  19/2 &   18.1 &23.8& 133.7\\
2& 6& 11/2      &  0.1 &  9& 9&  17/2 &   28.3 &28.2&  71.8\\
3& 7& 15/2      &  2.2 &  5& 8&  17/2 &    9.5 & 7.3& 114.5\\
3& 7& 15/2      &  2.2 &  8&10&  21/2 &   25.2 &22.9& 101.2\\
3& 7& 15/2      &  2.2 &  9&10&  21/2 &   29.3 &27.1& 108.7\\
\hline 
\end{tabular} 
\caption{The same as in Table \ref{table:dudr_1364sn_trans_2},
for the operator $dU/dr Y_{3M}$.}
\label{table:dudr_1364sn_trans_3}
\end{center}
\end{table}

\begin{table}
\begin{center}
\begin{tabular}{c|c|c|c}
\hline
 $l$ & $j$ & $E^{res}_{lj}$ [MeV] & $\Gamma^{res}_{lj}$ [MeV]  \\
\hline
 4  &9/2  &  -2.1 &  0.08   \\
 4  &7/2  &  -0.2 &  1.0   \\
 6  &11/2 &   0.2 &   0.04  \\
 7  &15/2 &   2.4 &    0.08  \\
 7  &13/2 &   8.7 &    2.5   \\
 8  &17/2 &   9.9 &    1.7  \\
 8  &15/2 &  20.3 &    8.0   \\
 9  &19/2 &  18.3 &   4.5   \\
10  &21/2 &  28.7 &   9.0  \\
\hline 
\end{tabular} 
\caption{Total and orbital angular momentum  ${lj}$, 
energies $E_{res}$  and widths
$\Gamma_{res}$ of the resonant neutron single particle states calculated extrapolating the HF potential of the  $^{1364}$Sn WS cell up to infinity. The continuum spectrum starts at 
$E_{cont}=$ -3.5 MeV.}  
\label{table:dudr_1364sn_reso}
\end{center}
\end{table}

From the analysis described above, we conclude that the collective response of the 
system is largely driven by the transitions involving the deeply bound
protons and the resonant neutron levels. However, the coupling with the
continuum shifts the energy of the peaks to lower energy and increases
considerably the width of the states, as well as the strength of the low-energy peaks.
We remark that the low-lying part of the spectrum would be modified by neutron
pair correlations, not taken into account in the present study,
although there are no  resonant single-particle 
states lying very close to  the Fermi energy ($E_F =$ 5.5 MeV), cf.
Table \ref{table:dudr_1364sn_reso}.

\clearpage

\subsubsection*{(b) The case of $^{498}$Zr}

\begin{figure*}[!t]
\centering
\includegraphics[width=0.36\textwidth]{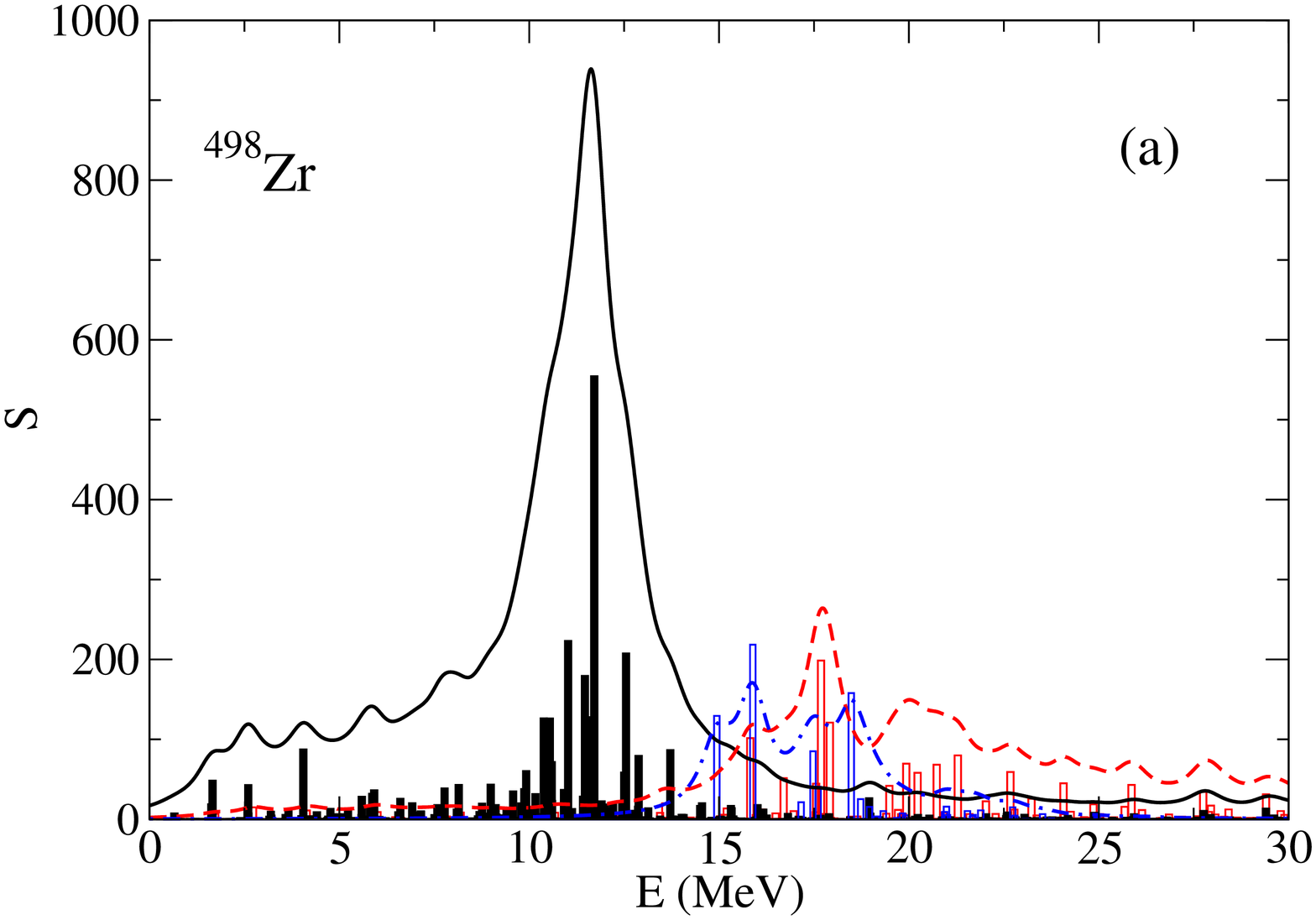}
\includegraphics[width=0.36\textwidth]{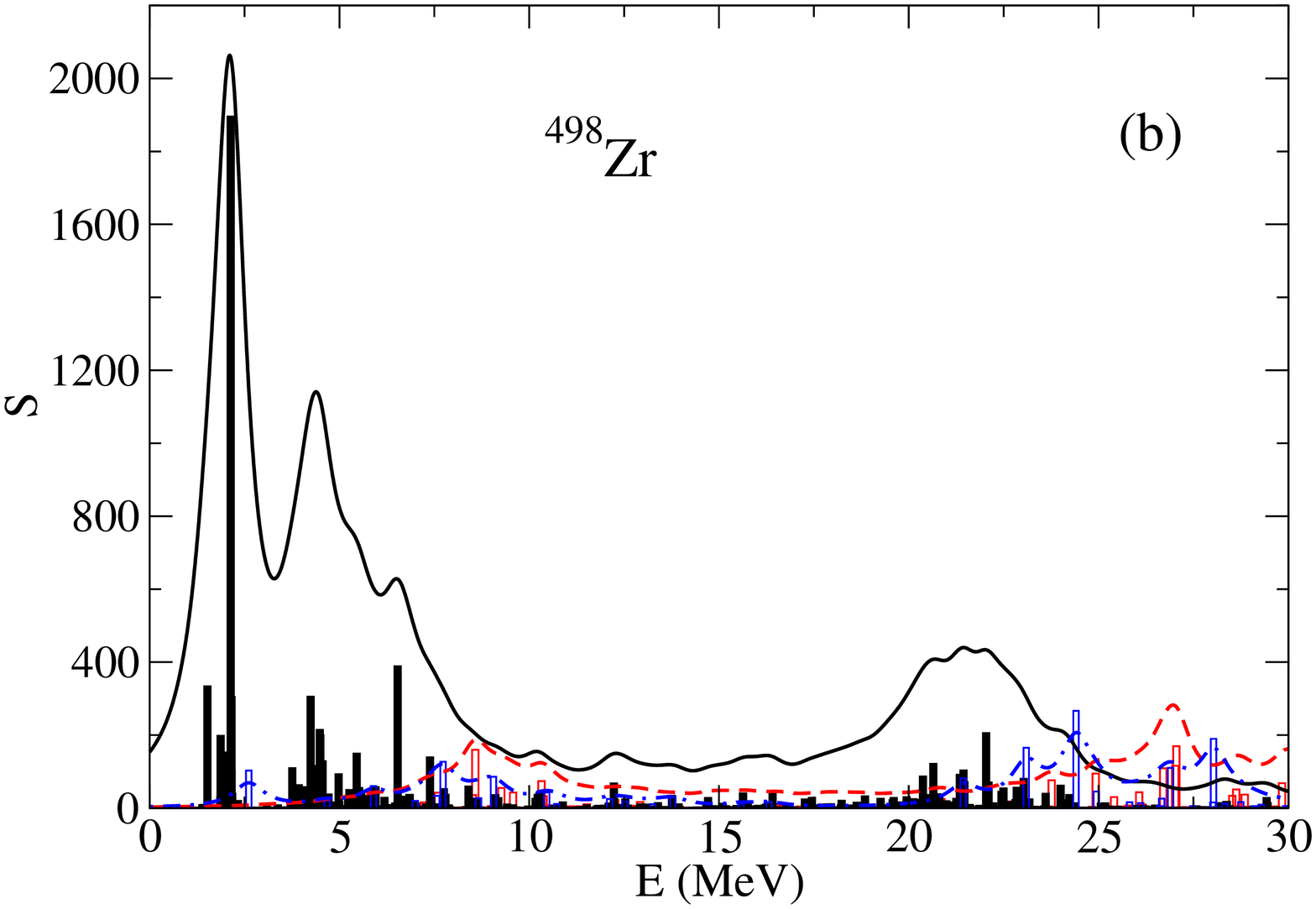}
\caption{ Quadrupole (a) and octupole (b) strength functions calculated in the $^{498}$Zr WS cell.
Black histograms refer to the discrete RPA response, while red and blue histograms refer to the discrete HF response for neutron and protons respectively (in units of MeV$^2$ fm$^{-2}$).
The solid curve  (in units of MeV fm$^{-2}$) 
is  obtained  by a convolution of the discrete RPA strength 
with a Lorentzian function having a FWHM
equal to 1 MeV. The dashed and dash-dotted curve are obtained convoluting 
the neutron and proton discrete HF strength respectively.}
\label{fig:498_response}
\end{figure*}

In this section we shall discuss the  results of a study  analogous to that 
carried out for $^{1364}$Sn, but this time for the
$^{498}$Zr  WS cell. 
The HF and RPA quadrupole and octupole strength functions
are shown in Fig. \ref{fig:498_response}. The main differences as 
compared to $^{1364}$Sn are found in the low-energy part of the response, where one 
notices the absence of the low-lying peak in the quadrupole
strength and the presence of a strong, very low-energy peak in the octupole response.
Both these features are related to the change occurring in the proton single-particle spectrum going from $Z=50$ to $Z=40$. In fact in $^{1364}$Sn  one finds a gap
of about 3 MeV between the last occupied proton orbit $ 1 g_{9/2}$ and the 
first unoccupied orbit $1 g_{7/2}$. Low-lying quadrupole transitions 
are then possible  between  the $ 1 g_{9/2}$  orbital 
and the $1 g_{7/2}$ or 2 $d_{5/2}$ orbitals. On the
other hand, the lowest octupole transitions   have an energy of
about 6 MeV ($1 \hbar \omega_p$). In $^{498}$Zr  the $ 1 g_{9/2}$  orbital is not occupied and 
low-energy octupole transitions are possible $( p_{3/2} \to g_{9/2}$ with an energy 
of 2.7 MeV), and only 2 $\hbar \omega_p$ quadrupole transitions are possible. 
The application of these results to the actual case of the inner crust 
is  model dependent. In fact, as we indicated above, detailed
calculations of the isotopic composition of the crust predict  different values for
the favoured proton number, typical values being 
$Z \approx 40$ or $Z \approx 50$. In Figs. \ref{fig:dudr_evolution}(e),(f) 
below we show the 
strength functions calculated for $^{506}$Sn ($Z=50$). As expected, in this case 
one recovers the low-lying peak in the quadrupole response while the low-lying peak
in the octupole response disappears.

Proton pairing correlations 
would lead to the partial occupation of the $ 1 g_{9/2}$ orbital, 
allowing the simultaneous presence of quadrupole and octupole low energy  transitions.
The low-lying part of the  spectrum might also be affected by neutron pairing correlations.

\begin{table}
\begin{center}
\begin{tabular}{cccc|cccc|cc}
\hline
$n_h$& $l_h$ & $j_h$ & $E_h$ &$n_p$&  $l_p$& $j_p$ & $E_p$ & $E_{ph}$ &
$T$[MeV$^2$fm$^{-2}$]\\ 
\hline
1& 4& 7/2       & -12.3 & 6&  6&  11/2 &    8.0 & 20.3  & 58.2\\
1& 4& 7/2       & -12.3 & 7&  6&  11/2 &    9.0 & 21.3  & 79.8\\
1& 5& 11/2      &  -7.6 & 7&  7&  15/2 &   10.1 & 17.7  &198.6\\
1& 5& 9/2       &  -1.3 & 10&  7&  13/2 &   18.6 & 19.9  & 69.7\\
1& 5& 9/2       &  -1.3 & 11&  7&  13/2 &   21.4 & 22.7  & 59.2\\
{2}& {6}& {13/2}      &   1.5 &  9&  8&  17/2 &   17.3 & 15.8  &101.4\\
{2}& {6}& {13/2}      &   1.5 &  8&  8&  17/2 &   19.4 & 17.9  &120.9\\
{2}&{6}&  {13/2}      &   1.5 &  8&  8&  17/2 &   22.2 & 20.7  & 68.1\\
\hline 
\end{tabular} 
\caption{List of the eight unperturbed neutron particle-hole transitions 
calculated in the $^{498}$Zr WS cell and 
associated with the largest transition strengths with the operator $dU/dr Y_{2M}$. 
In the first four columns we give the orbital angular momentum, the total
angular momentum and the energies  $l_{h},j_h$ and $E_h$ of the hole; in the 
next four columns, the  corresponding quantities $l_{p},j_p$ and $E_p$
for the particle. In the last two columns, we give the energy of the particle-hole jump 
$E_{ph} = E_p - E_h$ and its transition strength $T$. All energies are in MeV.}
\label{table:dudr_498zr_trans_2}
\end{center}
\end{table}

\begin{table}
\begin{center}
\begin{tabular}{cccc|cccc|cc}
\hline
$n_h$ & $l_h$ & $j_h$ & $E_h$ &$n_p$&  $l_p$& $j_p$ & $E_p$ & $E_{ph}$&  $T$[MeV$^2$fm$^{-2}$]\\ 
\hline
1& 5& 11/2      &   -7.6 & 9&  8&  17/2 &    17.3 &24.9&  94.6\\
1& 5&  9/2      &   -1.3 & 6&  6&  11/2 &    8.0 & 9.3&  55.1\\
1& 5&  9/2      &   -1.3 & 7&  6&  11/2 &    9.0 &10.3&  73.7\\
2& 3&  5/2      &   -0.5 & 6&  6&  11/2 &    8.0 & 8.5&  35.5\\
2& 3&  5/2      &   -0.5 & 7&  6&  11/2 &    9.0 & 9.5&  42.7\\
2& 6& 13/2      &    1.5 & 6&  7&  15/2 &    9.1 & 7.6&  42.0\\
2& 6& 13/2      &    1.5 & 8&  7&  15/2 &   10.1 & 8.6& 159.5\\
2& 6& 13/2      &    1.5 &11&  9&  19/2 &   25.2 &23.7&  75.8\\
\hline 
\end{tabular} 
\caption{The same as in Table \ref{table:dudr_498zr_trans_2},
for the operator $dU/dr Y_{3M}$.}
\label{table:dudr_498zr_trans_3}
\end{center}
\end{table}

\begin{table}
\begin{center}
\begin{tabular}{c|c|c|c}
\hline
 $l$ & $j$ & $E_{res}$ [MeV] & $\Gamma_{res}$ [MeV]  \\
\hline
 4  &9/2  &   4.7   & 2.0   \\
 6  &13/2 &   1.6   & 5 $\times 10^{-3}$   \\
 6  &11/2 &   8.8   & 1.2   \\
 7  &15/2 &   10.1  & 1.0   \\
 7  &13/2 &   21.0  & 6.5  \\
 8  &17/2 &   19.0  & 3.7   \\
 9  &19/2 &   30.3  & 8.5  \\
\hline 
\end{tabular} 
\caption{Total and orbital angular momentum  ${lj}$, 
energies $E_{res}$  and widths
$\Gamma_{res}$ of the resonant neutron single particle states calculated in the
$^{498}$Zr WS cell. The continuum spectrum starts at $E_{cont}= $ -0.5 MeV.}  
\label{table:dudr_498zr_reso}
\end{center}
\end{table}

\begin{figure*}[!h]
\begin{center}
\includegraphics[width=0.36\textwidth]{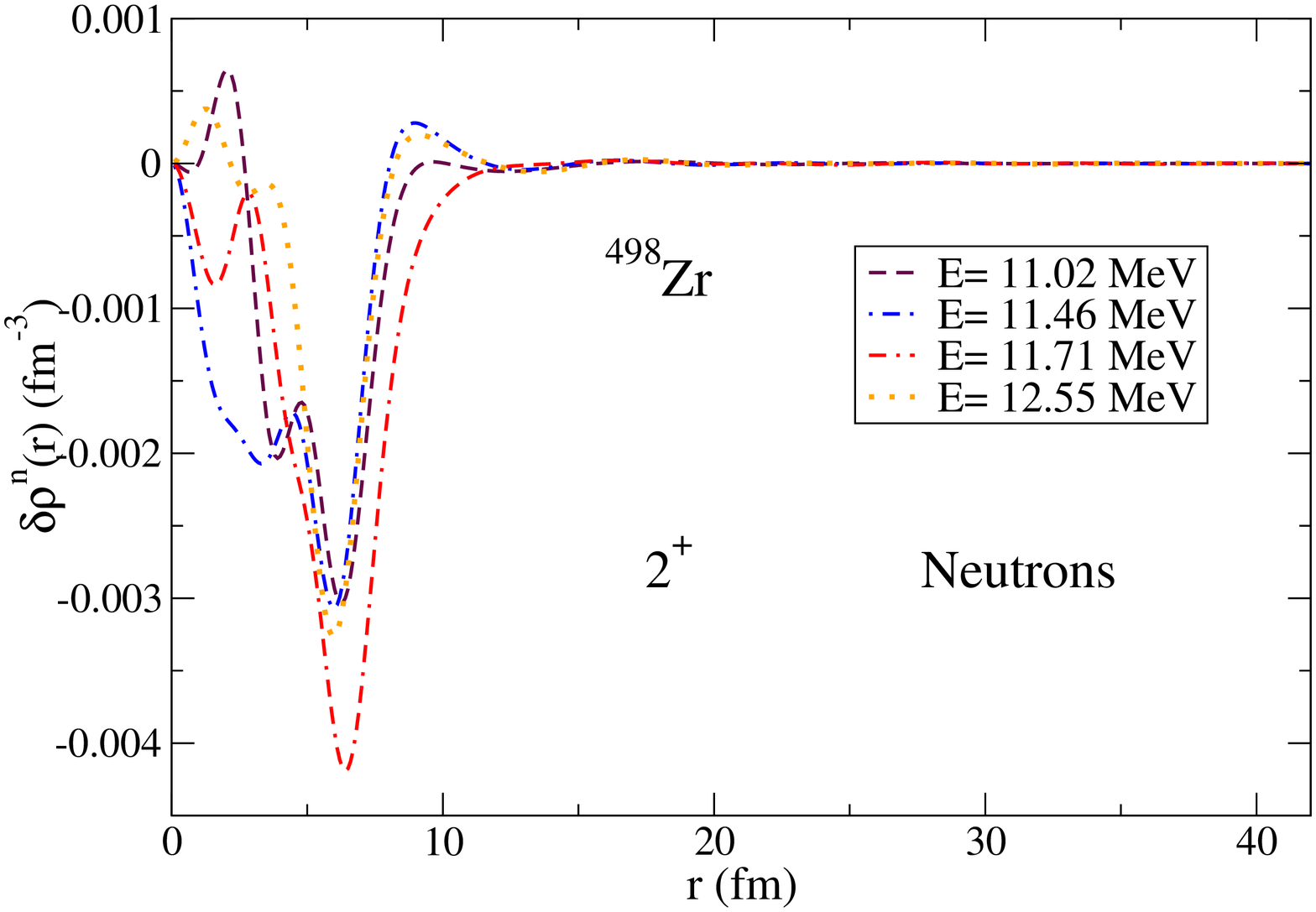}
\includegraphics[width=0.36\textwidth]{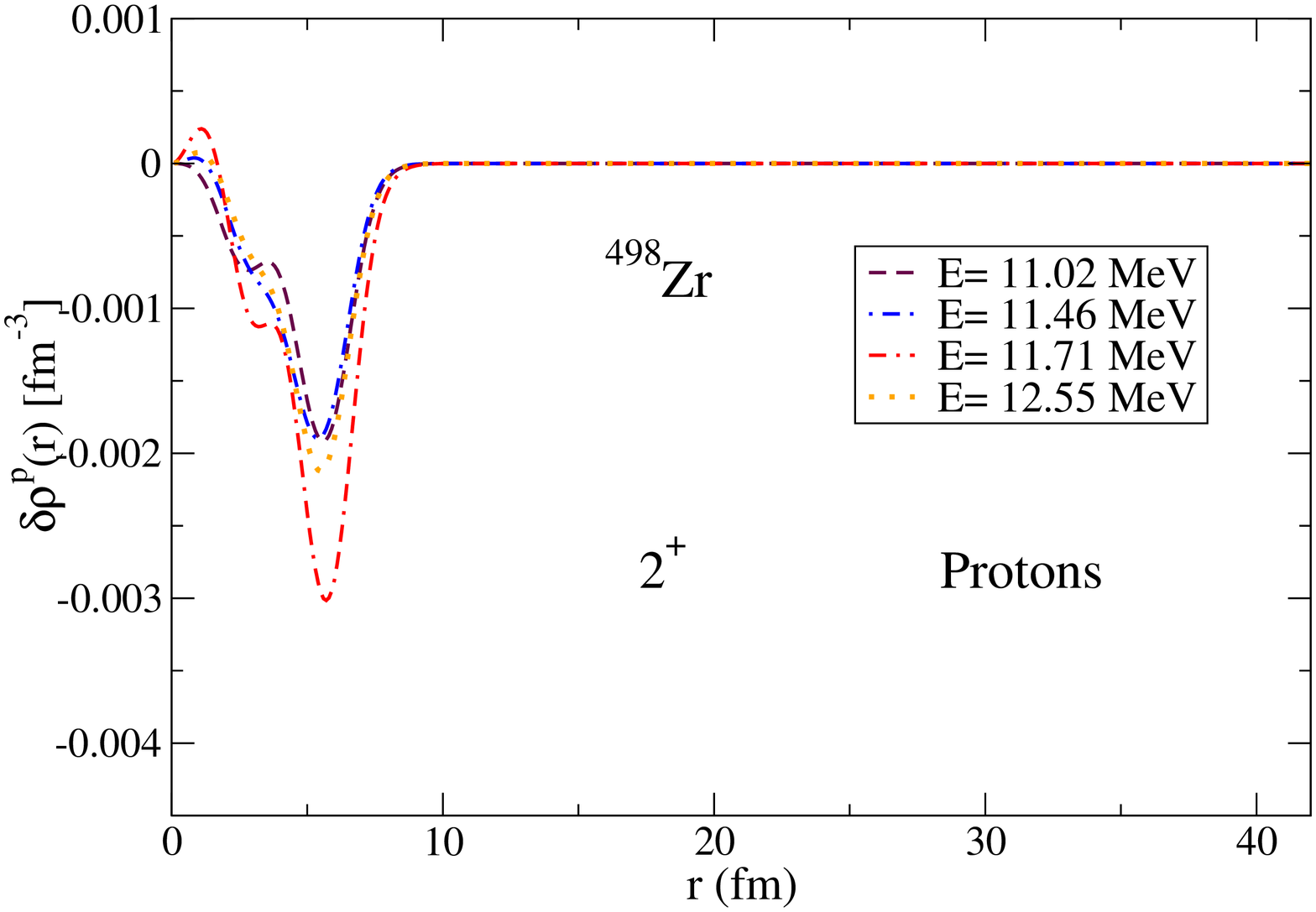}
\end{center}
\caption{Neutron (left panel) and proton (right) transition densities associated with the four strongest 
RPA transitions calculated in the response  
to the operator $dU/dr Y_{2M}$ for the $^{498}$Zr  WS cell, shown in Fig. \ref{fig:498_response}(a). 
The energies of the transitions are listed in the legends of the figures. }
\label{fig:trans_498zr}
\end{figure*}

The analysis of the neutron transitions leads to the same conclusions
presented above in the case of $^{1364}$Sn.  
On the one hand, the transition densities associated with 
the strongest quadrupole phonons
(cf. Fig. \ref{fig:trans_498zr}) have a very clear surface character. 
On the other hand,  the strongest neutron 
particle-hole transitions (listed in Tables \ref{table:dudr_498zr_trans_2}
and \ref{table:dudr_498zr_trans_3}) are again  associated with the 
resonant single-particle states identified  by the phase shifts and  
listed in Table \ref{table:dudr_498zr_reso}.  

\begin{figure*}[!h]
\begin{center}
  \includegraphics[width=0.36\textwidth]{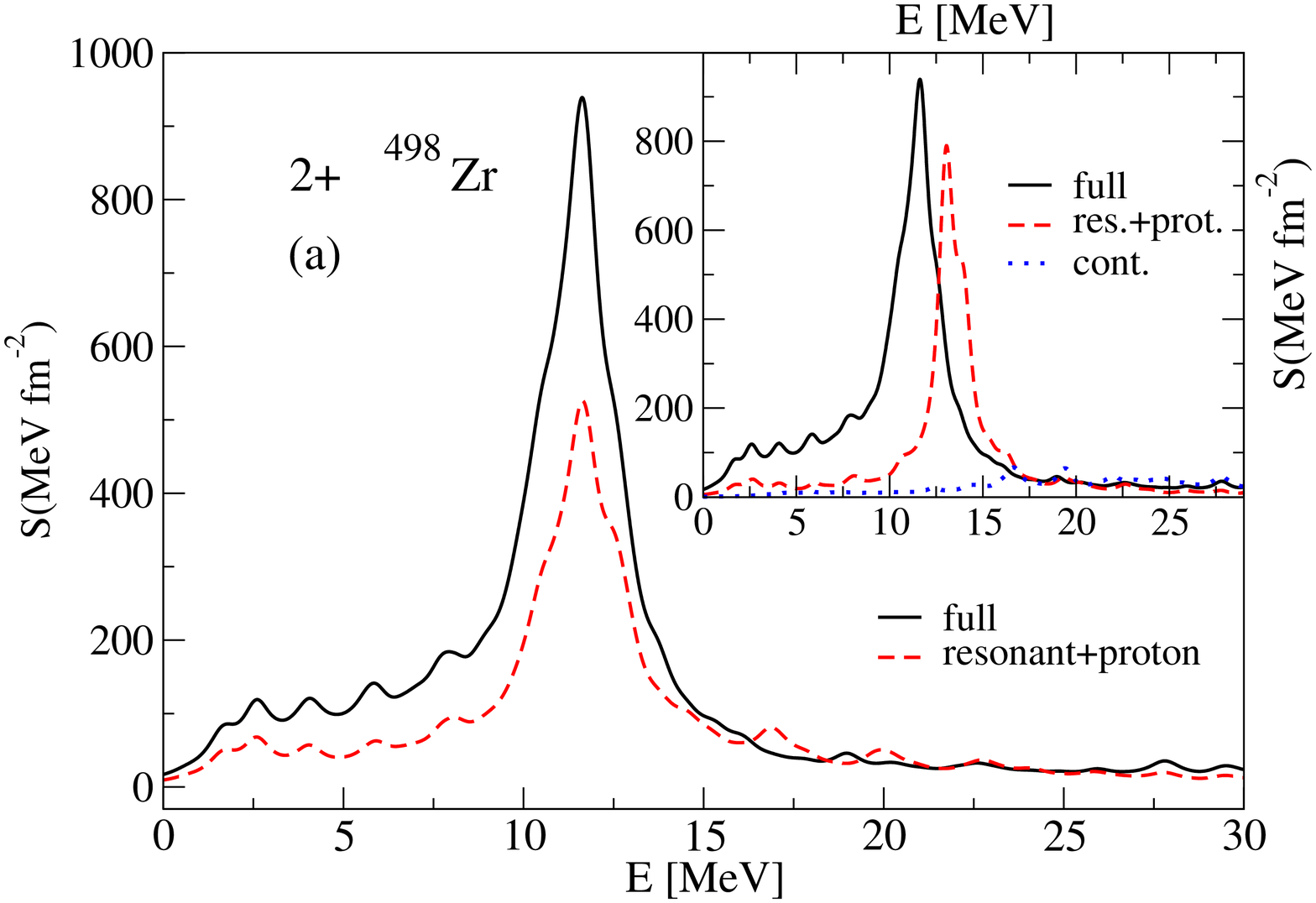}
\includegraphics[width=0.36\textwidth]{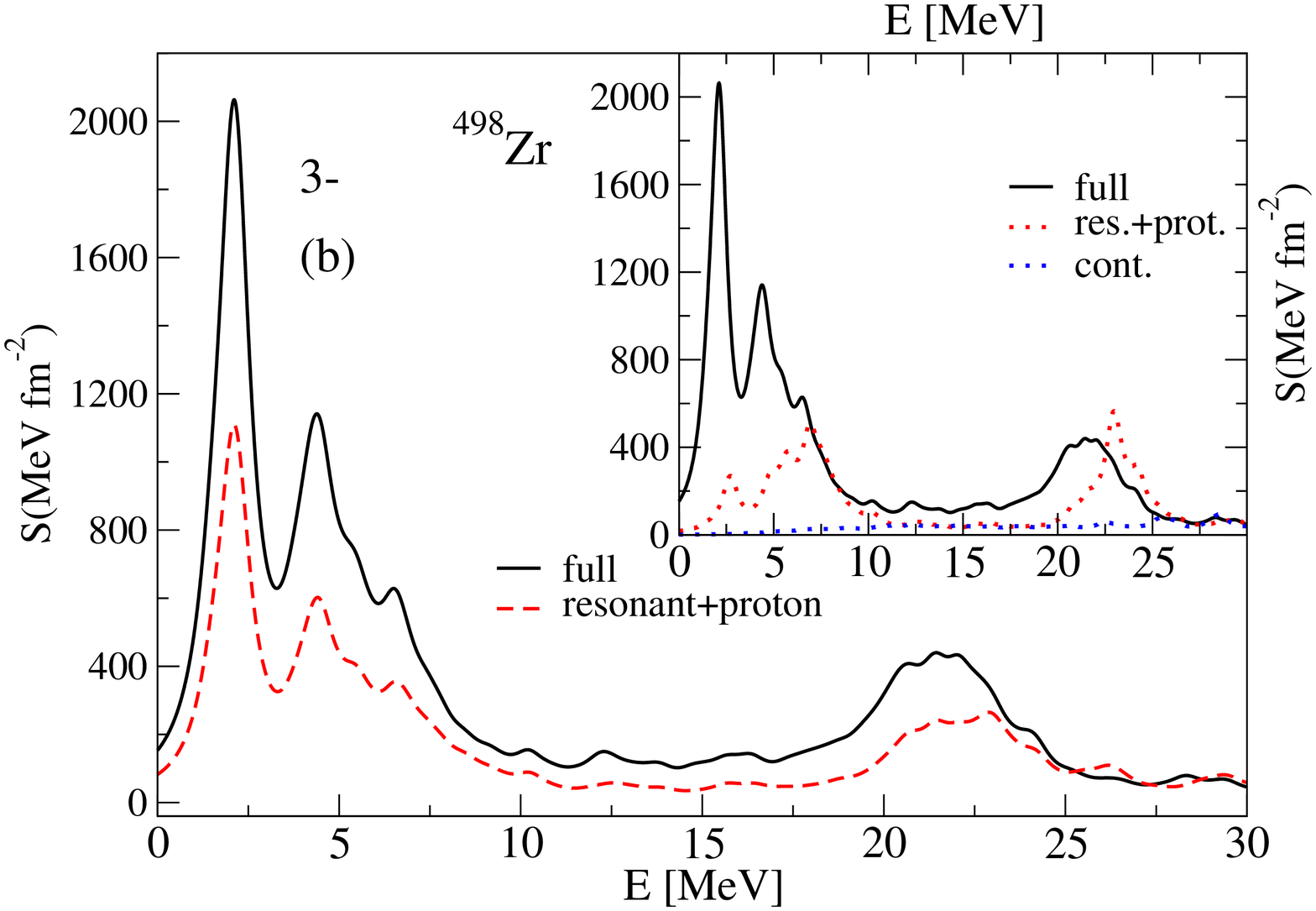}
\end{center}
\caption{(a) The RPA quadrupole strength function calculated for the $^{498}$Zr  WS cell
(solid line, cf. Fig.\ref{fig:498_response}(a)), is compared with the strength 
obtained including only the amplitudes associated with proton and  resonant neutron 
particle-hole transitions (dashed line, cf. text for details).
In the inset, the RPA  quadrupole strength function is instead
compared with the RPA strength
calculated using a restricted particle-hole basis limited either to proton states and to 
resonant neutron states (dashed line) or to continuum neutron states (dotted line).
(b) The same as (a), for the operator $dU/dr Y_{3M}$.}
\label{fig:dudr_498zr_restricted_23}
\end{figure*}

Furthermore, the analysis of the strength functions 
(cf.  Fig. \ref{fig:dudr_498zr_restricted_23}) shows that the shape of the response
is well reproduced including only the amplitudes over proton and resonant
states. Similarly to $^{1364}$Sn, also 
in $^{498}$Zr a RPA calculation including only proton and resonant states
produces a narrow collective peak in the quadrupole strength, close to 13 MeV 
(cf. the dashed curve in the inset of Fig. \ref{fig:dudr_498zr_restricted_23}). Compared to the $^{1364}$Sn, however, the coupling to the neutron  continuum produces a smaller shift of the peak  and a smaller increase of the width, due to the reduced
density of the system.

\clearpage

\subsubsection* {(c) Evolution of the response with the neutron number}

Having seen that the resonant single-particle states largely determine the
$2^+$ and $3^-$ response to the $dU/dr Y_{LM}$ operator,
one may  expect that the general features of the 
response to the $dU/dr$ operator should evolve in a reasonably  continuous 
fashion going from the atomic nucleus into the inner crust.
In Fig. \ref{fig:dudr_shell_evolution}
we compare in a schematic way the quadrupole and octupole
neutron particle-hole transitions calculated 
$^{1364}$Sn and $^{498}$Zr with those calculated in the
closed-shell, neutron-rich nucleus $^{132}$Sn. The horizontal lines
indicate the energies of the neutron bound levels or of the resonant levels
(cf.  Tables  \ref{table:dudr_1364sn_reso} and 
\ref{table:dudr_498zr_reso} for $^{1364}$Sn and $^{498}$Zr). The solid and 
dashed lines  indicate the strongest  quadrupole and octupole 
transitions connecting bound or  resonant levels (cf. Tables 
\ref{table:dudr_1364sn_trans_2} and \ref{table:dudr_1364sn_trans_3} 
for $^{1364}$Sn and Tables \ref{table:dudr_498zr_trans_2}
and \ref{table:dudr_498zr_trans_3} for $^{498}$Zr). 
One can clearly recognize that the shell structure plays
a similar role  in the three cases, leading to unperturbed 
particle-hole transitions characterized by an energy $2 \hbar \omega_n$
(for quadrupole transitions) or $ 1 \hbar \omega_n$ and $3 \hbar \omega_n$
(for octupole transitions).

\begin{figure*}[!hb]
\centering
  \includegraphics[width=0.36\textwidth]{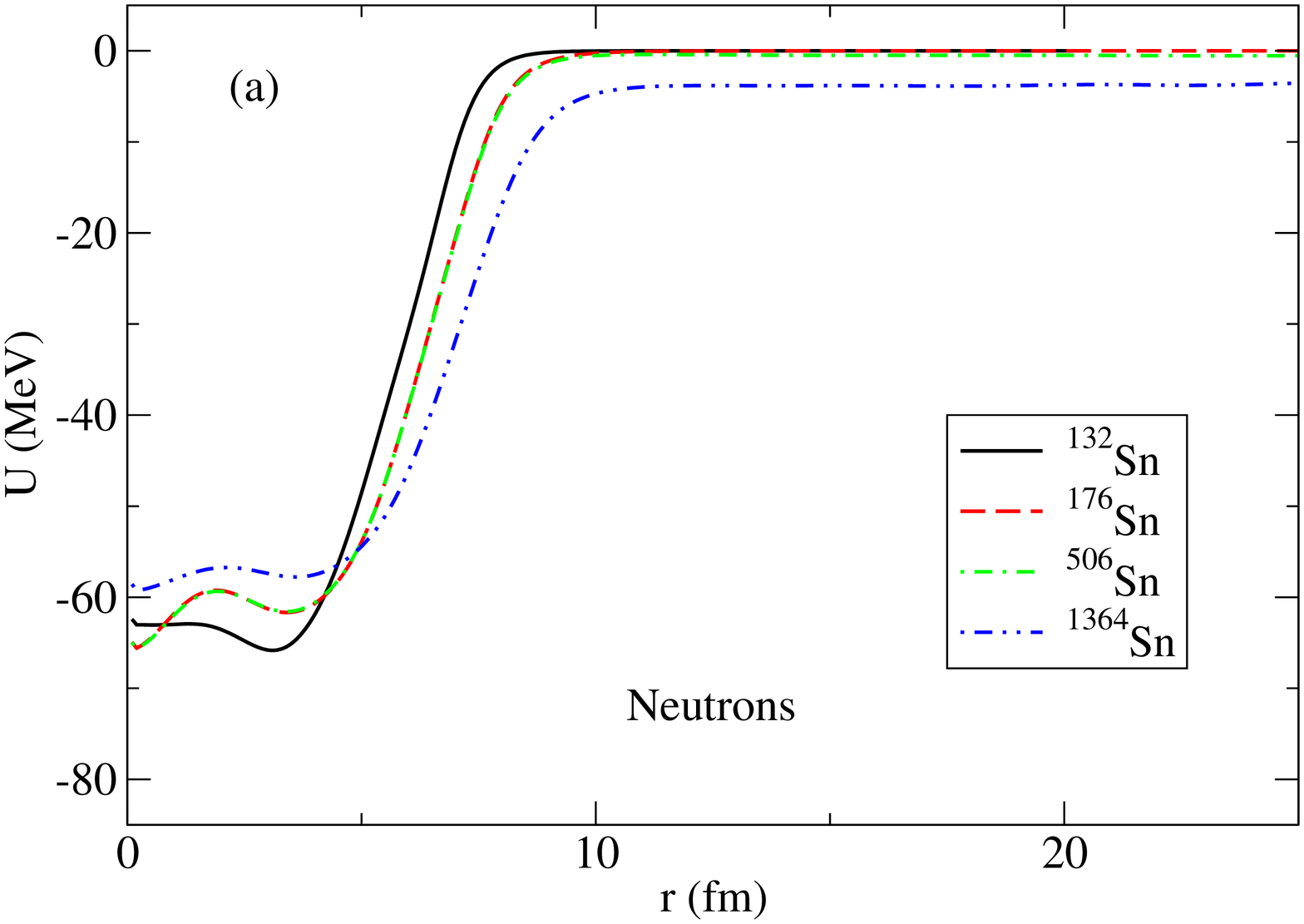}
\hspace{1cm}
 \includegraphics[width=0.36\textwidth]{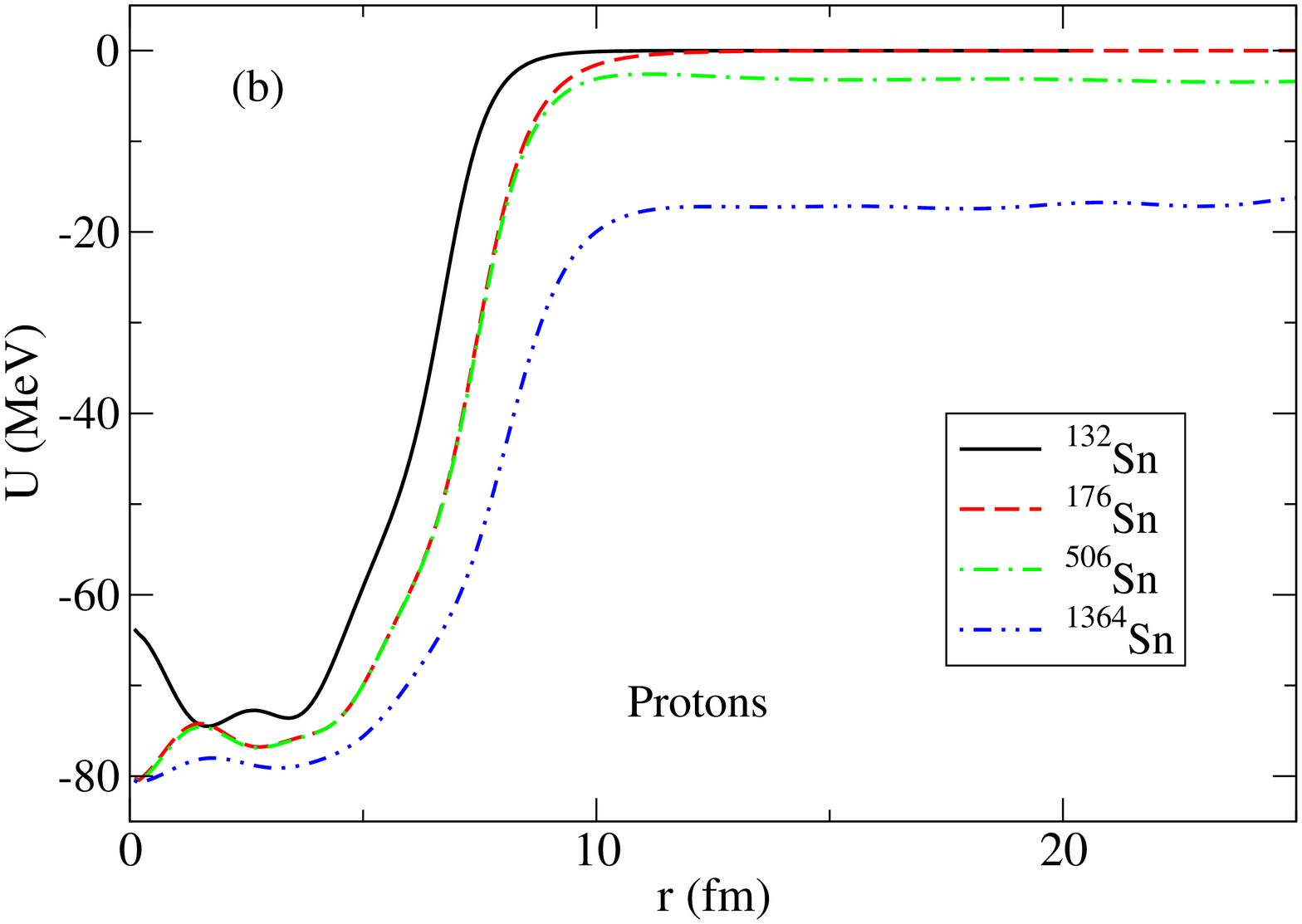}\\
\caption{Neutron (a) and proton (b) mean-field potentials
calculated for the nuclei $^{132}$Sn  and $^{176}$Sn and for the 
WS cells $^{506}$Sn  and $^{1364}$Sn.  }
\label{pot_evolution}
\end{figure*}

\begin{figure*}[!h]
\centering
  \includegraphics[width=0.36\textwidth]{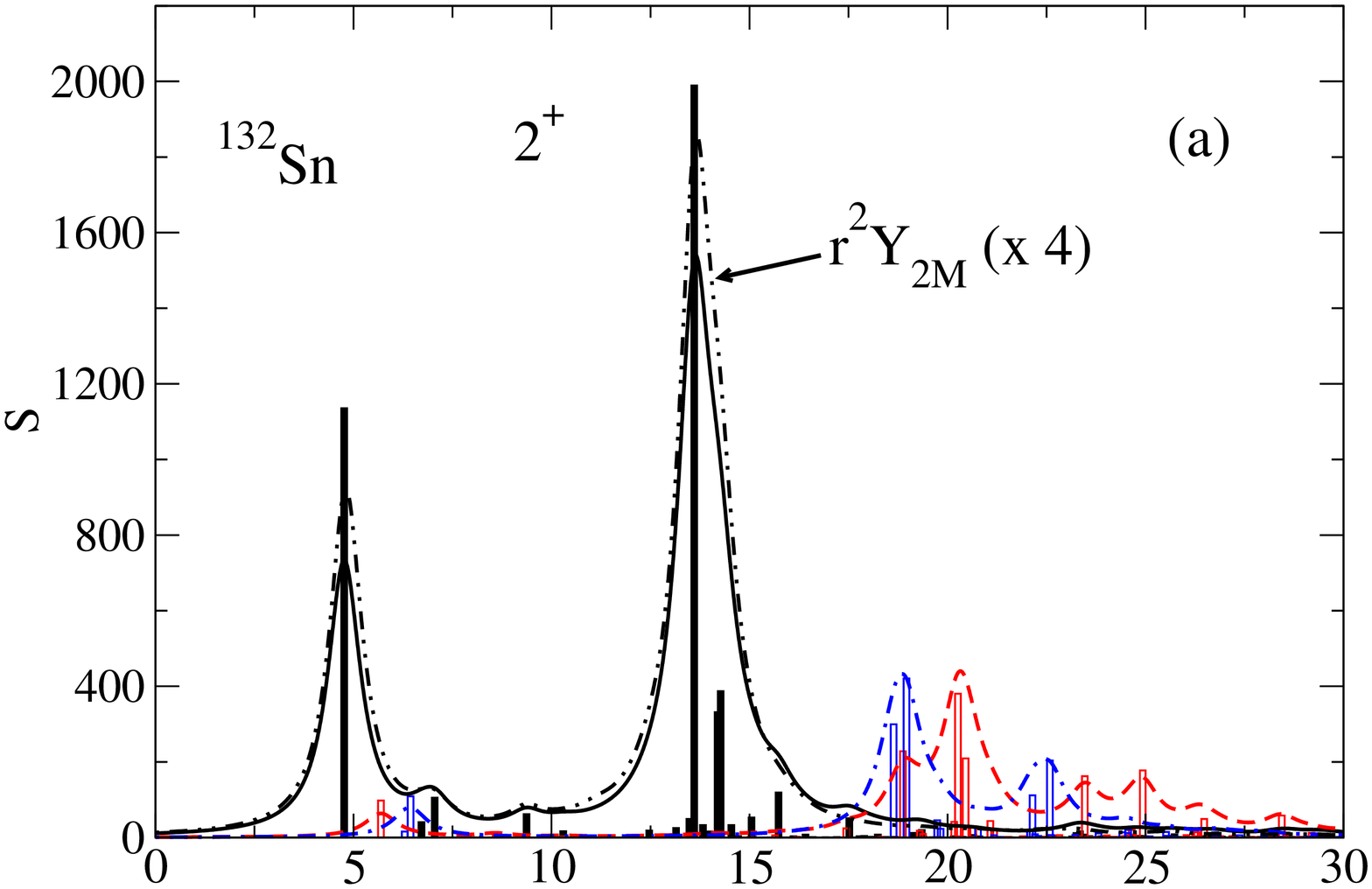}
 \includegraphics[width=0.36\textwidth]{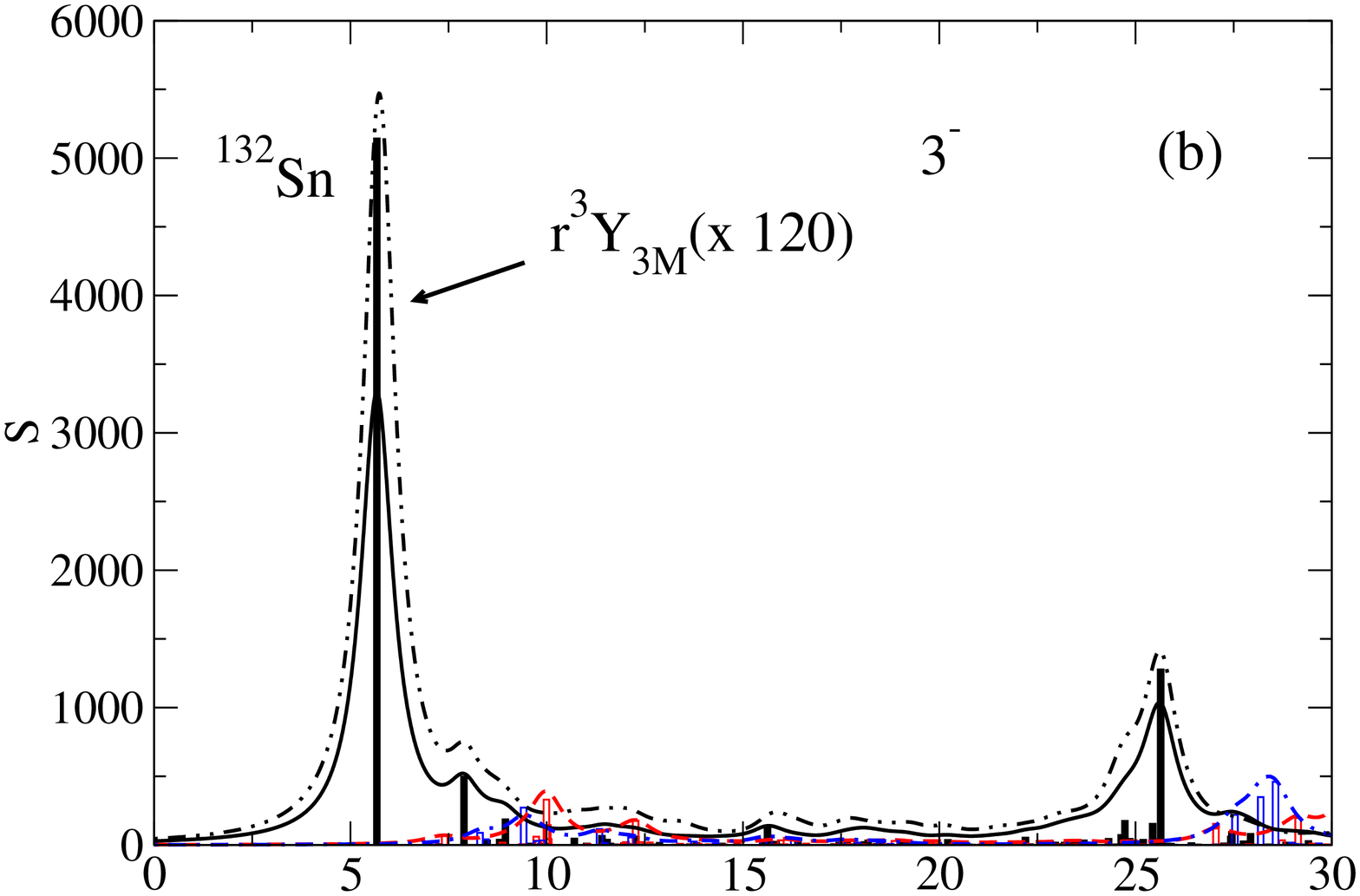}\\
\vspace{8mm}
  \includegraphics[width=0.36\textwidth]{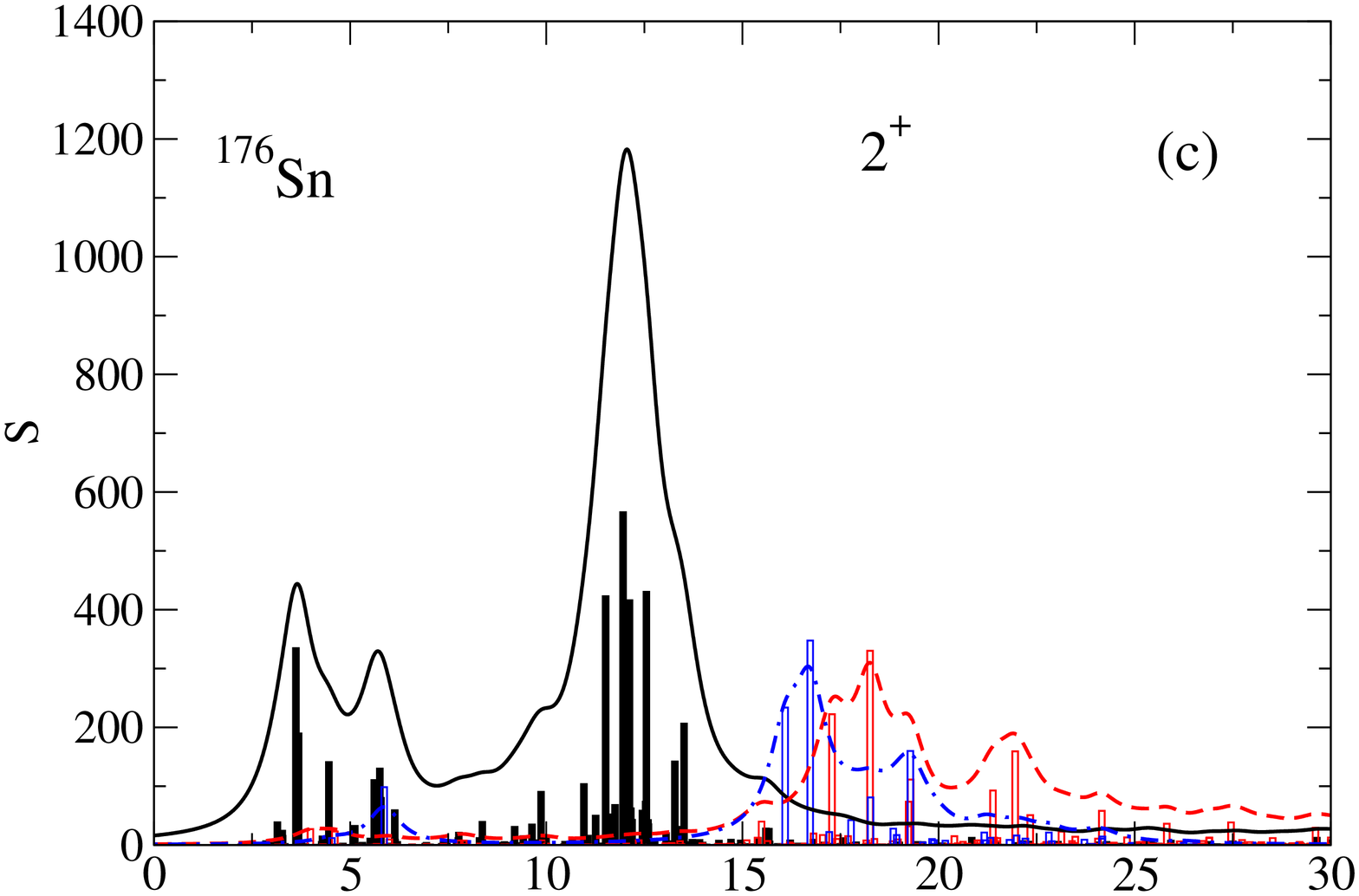}
 \includegraphics[width=0.36\textwidth]{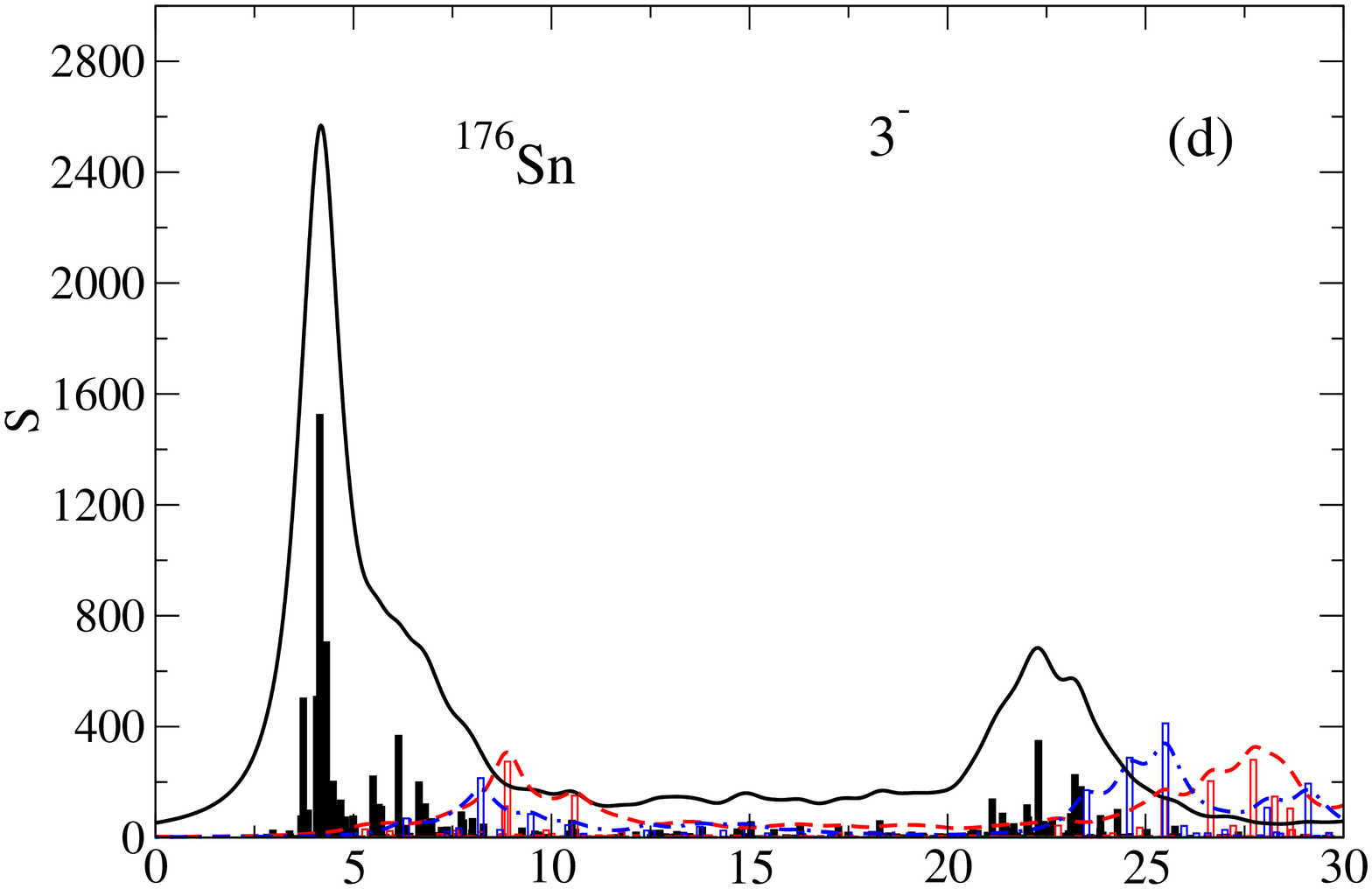}\\
\vspace{8mm}
   \includegraphics[width=0.36\textwidth]{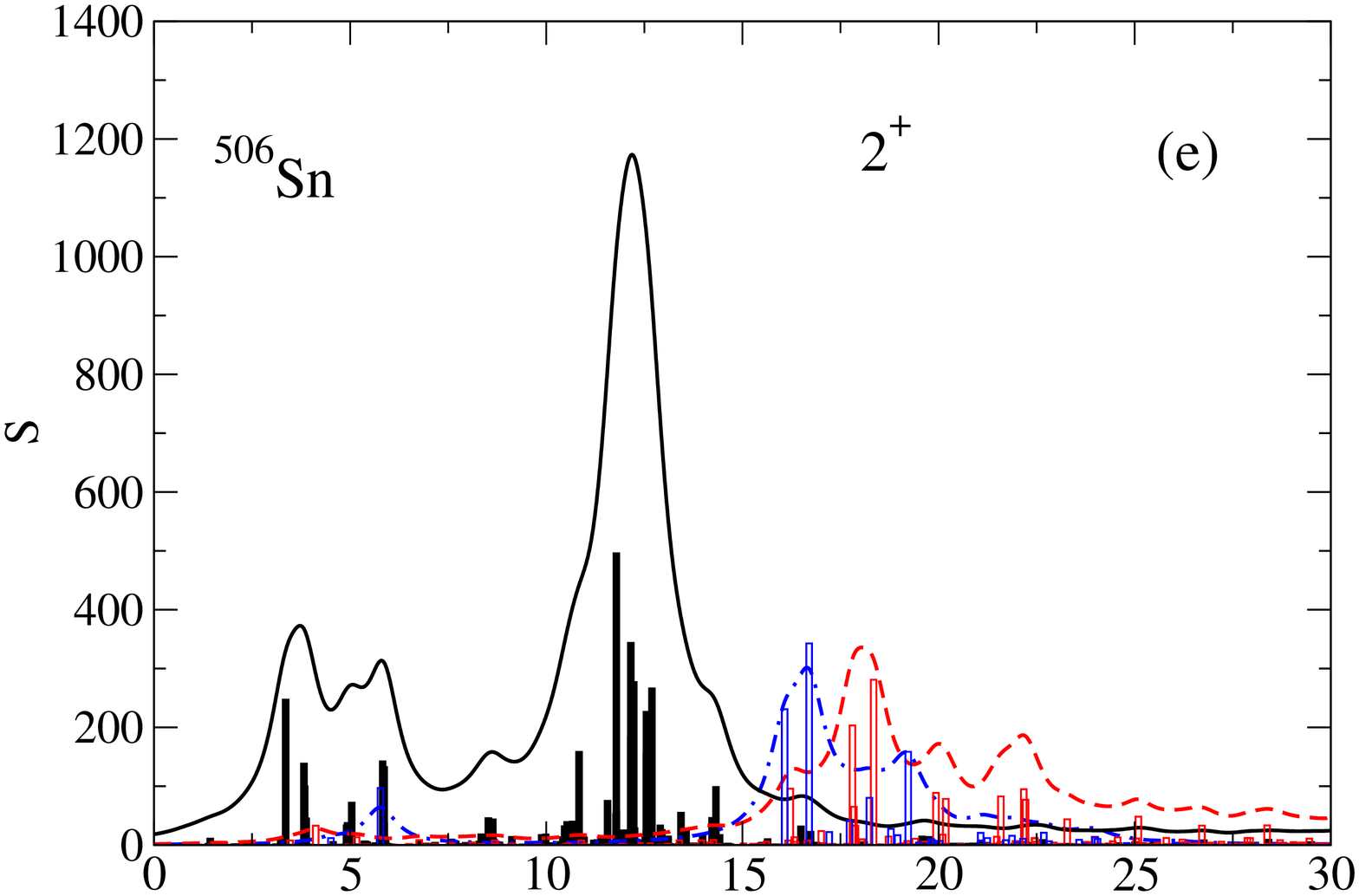}
 \includegraphics[width=0.36\textwidth]{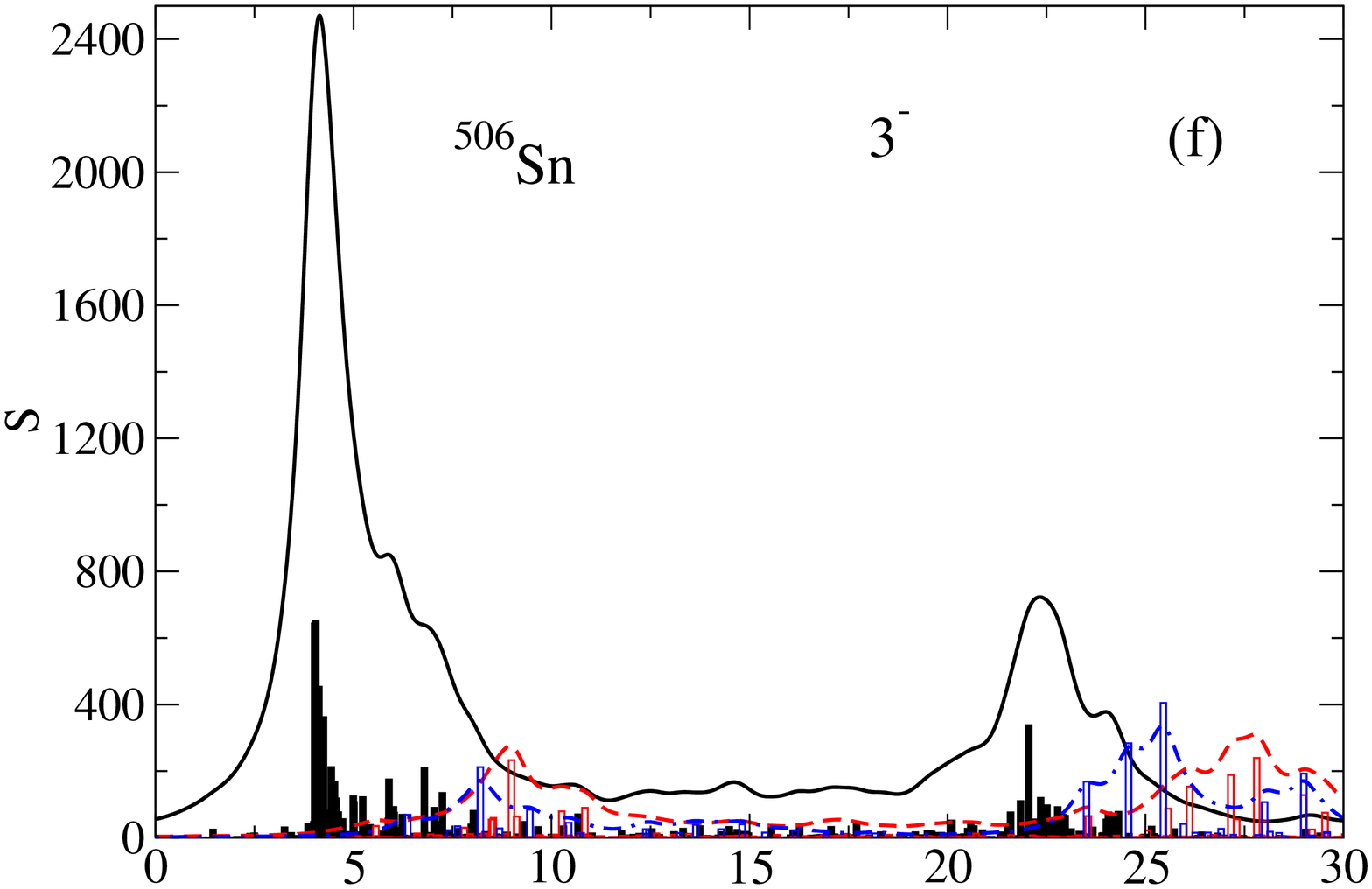}\\
\vspace{8mm}
  \includegraphics[width=0.36\textwidth]{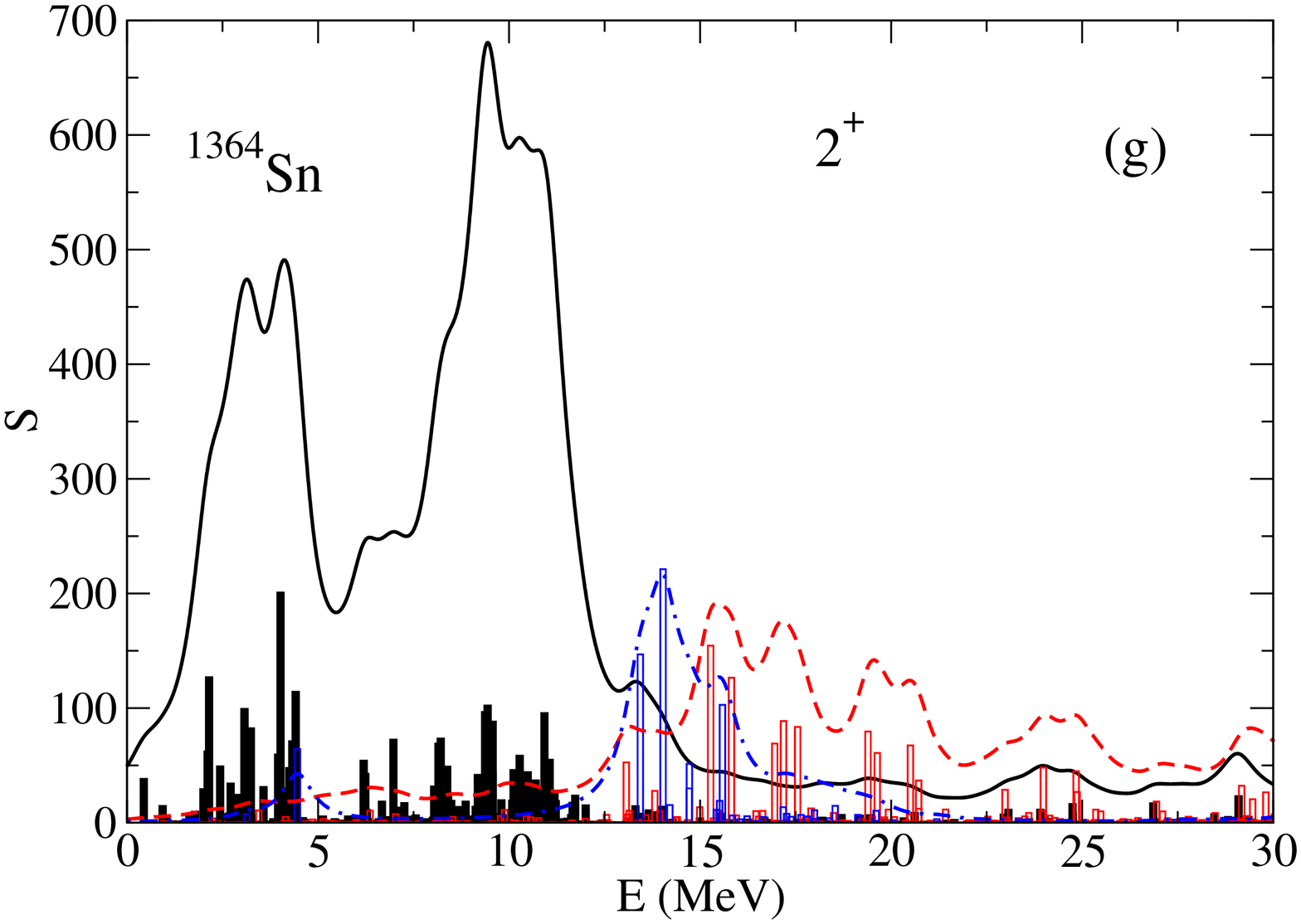}
 \includegraphics[width=0.36\textwidth]{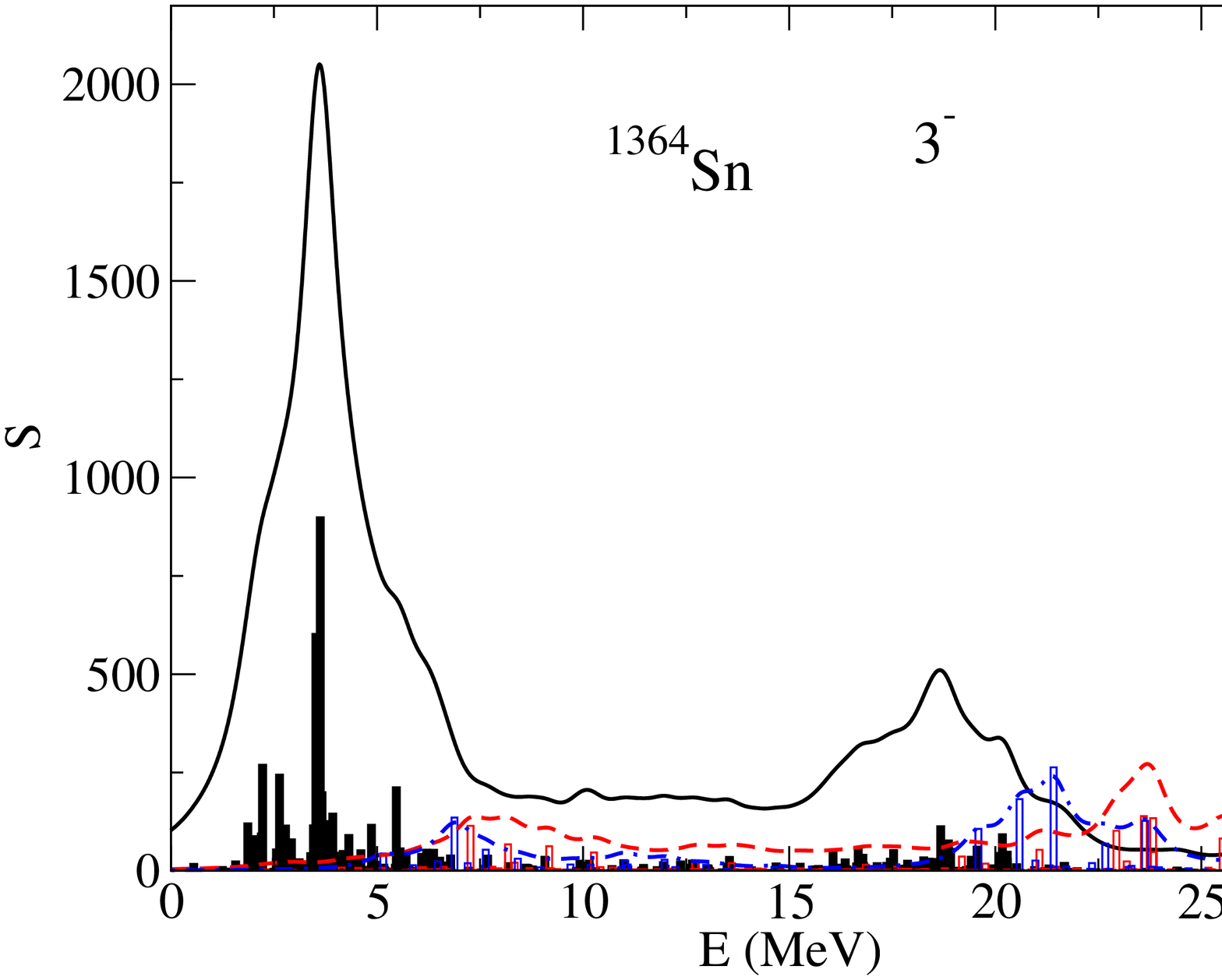}
\caption{Evolution of the quadrupole and octupole response going from
the neutron-rich nucleus $^{132}$Sn to the drip line and into the inner crust.  
We show the strength functions of the operators $dU/dr Y_{2M}$ 
and $dU/dr Y_{3M}$ calculated in $^{132}$Sn (panels 
(a) and (b)) ,$^{176}$Sn (panels 
(c) and (d)), $^{506}$Sn (panels 
(e) and (f))  and in $^{1364}$Sn (panels (g) and (h),
already shown in Fig. \ref{fig:1364_response}). 
 In all the panels, the black histograms
refer to the discrete RPA response while the red and blue histograms 
refer to the discrete HF response for neutrons and protons 
respectively   (in units of MeV$^2$ fm$^{-2}$). 
The red dashed, blue dash-dotted and solid curves (in units of MeV fm$^{-2}$) 
are obtained convoluting the  neutron and proton HF responses 
and the RPA response with a 
Lorentzian function having a FHWM equal to 1 MeV.  
In panels (a)  and (b) we also show the quadrupole and octupole responses
associated with the operators $r^2 Y_{2M}$ (in units of fm$^{-4}$ and multiplied by 4)
and $r^3 Y_{3M}$ in $^{132}$Sn (in units of fm$^{-6}$ and multiplied by 120).}
\label{fig:dudr_evolution}
\end{figure*}

\begin{figure*}[!h]
\begin{center}
\vspace{20mm}
  \includegraphics[width=0.5\textwidth]{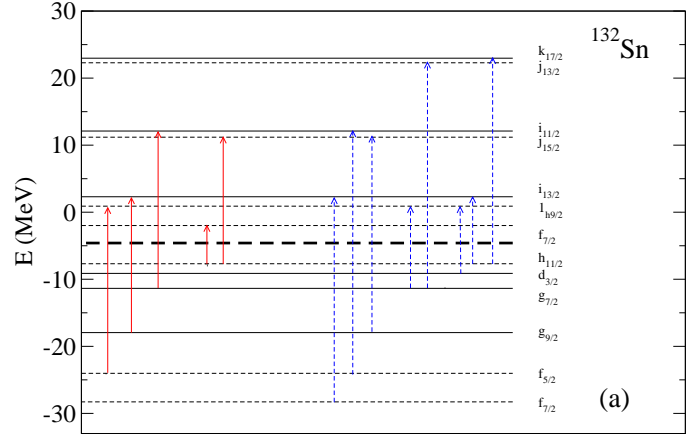}\\
\vspace{20mm}
\includegraphics[width=0.5\textwidth]{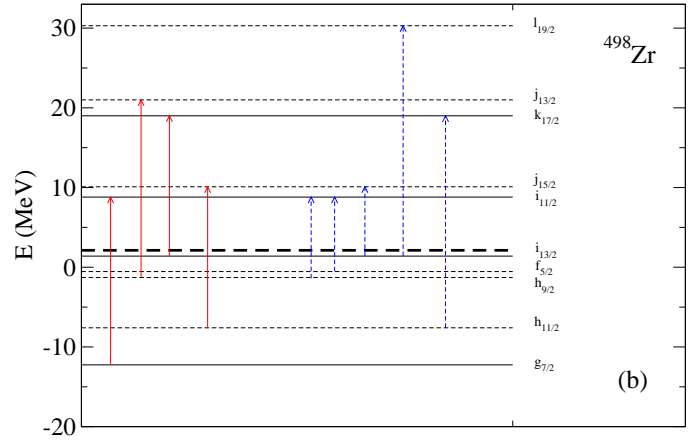}\\
\vspace{20mm}
\includegraphics[width=0.5\textwidth]{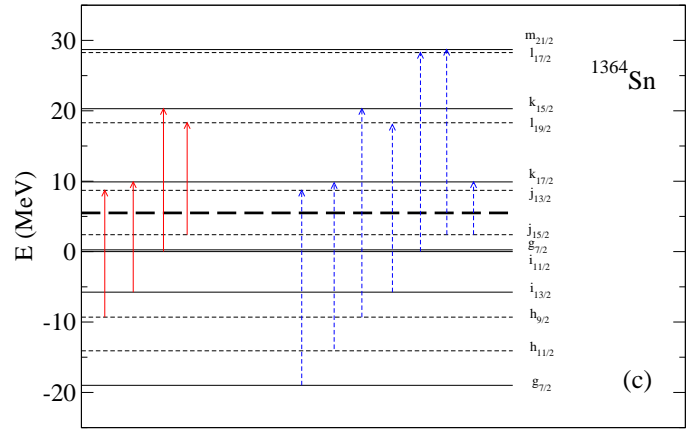}
\end{center}
\caption{The strongest neutron particle-hole transitions in $^{132}$Sn 
(a), in $^{498}$Zr (b) and in $^{1364}$Sn (c), see text for details.
Single-particle levels of even and odd parity are drawn by solid and dashed lines. Solid
(red) and dashed (blue) arrows refer to $2^+$ and to $3^-$ transitions. The Fermi energy is represented by the thick dashed line.}
\label{fig:dudr_shell_evolution}
\end{figure*}

In Fig. \ref{fig:dudr_evolution} we compare the 
the $2^+$ and $3^-$ strength functions in the inner crust 
with those of atomic nuclei. 
We keep the same number of protons ($Z=50$) and increase the number of neutrons,
going from the neutron-rich, closed-shell nucleus $^{132}$Sn
(panels (a),(b)) to the 
drip-line, closed-shell  nucleus  $^{176}$Sn ((c),(d)) and then to 
the WS cell $^{506}$Sn ((e),(f)) and finally to $^{1364}$Sn
((g),(h)). The calculation of $^{506}$Sn
is very similar to that discussed above for $^{498}$Zr, the only difference
being the number of protons. 
The proton and neutron  mean-field potentials are compared
in Fig. \ref{pot_evolution}.

One can see from Fig. \ref{fig:dudr_evolution}  that the main 
features of the giant resonances  are the same in the four cases. 
The centroids of
the main peaks, associated with 
$2 \hbar \omega$ transitions in the quadrupole response and 
with $1 \hbar \omega$ and $3 \hbar \omega$ in the octupole
response, are lowered going from $^{132}$Sn to $^{1364}$Sn.
This is in part due to the shift of the unperturbed response, associated with 
the larger radii of the potentials, particularly in the case of protons,
and in part is caused by  the coupling with the neutron
continuum, as we discussed above. The coupling also increases the
width of the peaks. A more complete study should  include the
coupling to $2p-2h$ configurations.  

In Fig. \ref{fig:dudr_evolution}(a),(b) we show also the strength functions
obtained 
with the operators $r^2 Y_{2M}$ and $r^3 Y_{3M}$ in the case of $^{132}$Sn.
Their shape essentially  coincide with those calculated making use of
the operator $dU/dr Y_{LM}$;
this is not the case for the calculations performed in the inner crust,
as discussed in Appendix B. 

\clearpage

\section{Conclusions}

As Negele and Vautherin noticed in their seminal paper,  'the degree 
to which the nuclei in the free neutron regime resemble ordinary nuclei' is
striking, and 'this similarity is also manifested  in the behaviour of 
the single-particle energies' \cite{Neg.Vau:73}. In this work we have
shown that the persistence of the shell structure influences the linear response
of the Wigner-Seitz cell in the inner crust  to a large extent.
We have found it useful to  study the response to the operator $dU/dr Y_{LM}$, 
which acts as a filter, enabling one  to focus on the effects associated with the nuclear surface.
Due to the existence of resonant states with rather narrow widths, the main features of 
quadrupole and octupole giant resonances are similar to those of ordinary
atomic nuclei. However, the interaction of the bound nucleons
with the neutron sea, has an important effect, leading to a decrease of the energy of the peaks, 
and to an increase of their width.

The existence of vibrational modes associated with the nuclear surface can have
important consequences on the neutron superfluidity of the system in the $^1S_0$ channel. 
While  medium effects tend to suppress pairing correlations in neutron matter, 
being dominated by spin fluctuations,  the surface fluctuations studied in this paper 
lead to an attractive contribution. Even if the global effects may not be large, 
due to small volume occupied by the nucleus, they can affect the spatial dependence
of the pairing gap, and have important consequences on vortex pinning \cite{simone,vortexnoi1,vortexnoi2}.
In this respect, it would be also important to examine the coupling of neutrons 
with lattice vibrations  \cite{bulgac,reddy}.

\section{Acknowledgment}
This work was supported in part by 
the NSF under Grant 0835543, the Natural Sciences and Engineering Research Council
of Canada (NSERC). TRIUMF receives
funding via a contribution through the National Research Council Canada.
This work was supported by the Academy of Finland and University of Jyv\"askyl\"a within the FIDIPRO program. 
F.B. acknowledges partial support
from the Spanish Education and Science Ministry.

\section{Appendix A}\label{app:a}

In this Appendix we shall compare the results of numerical
calculations of the linear response 
in a WS cell without protons with analytical
results which are available in uniform matter
in the case of Skyrme interactions \cite{Gar.ea:92}. 

We shall consider the excitation of either density or spin modes,
computing the response function  to the  external fields 
\begin{equation}
V_{ext}(\vec{r})=e^{i\vec{q}\cdot\vec{r}}\Theta^{(\alpha)},
\label{external}
\end{equation}
where $\alpha=0,1$ distinguishes the two possible channels $S=0$ and $S=1$:
$\Theta^{0}$ = 1, $\Theta^{1}=\sigma_z$ according to whether density 
or spin modes are considered.

In neutron matter, the RPA strength function can be written as
\begin{equation}
  S^{(\alpha)}(q,E)=- \frac{1}{\pi} Im \Pi^{(\alpha)}(q,E) = \nonumber
\end{equation}
\begin{equation}
- \frac{1}{\pi} \int d^3r d^3r' V_{ext}(\vec{r}) V_{ext}(\vec{r}') Im\Pi(\vec{r},\vec{r}',E),	
\label{NS comp: eq 8}
\end{equation}
where $\Pi(\vec{r},\vec{r}',E)$ is the polarization function \cite{Ber.Bro:1994}.

The RPA strength function calculated in neutron matter with 
a Skyrme effective interaction reads
\begin{eqnarray}\label{eq:cap1calc_RPA130}
&{\Pi^{(\alpha)}(\nu,k)}= 2 \Pi_{0} [ 1-W_{1}^{(\alpha)}\Pi_{0}   \nonumber \\
&-2W_{2}^{(\alpha)}k_F^2 \left(k^{2}-\frac{\nu^{2}}{1-(m_k k_{f}^{3}/3\pi^{2})W_{2}^{(\alpha)}}\right)\Pi_{0}  \nonumber \\
&+\left(W_{2}^{(\alpha)}k_{f}^{2}\right)^{2} (\Pi_{2}^{2}-\Pi_{0}\Pi_{4}+4k^{2}\nu^{2}\Pi_{0}^{2} 
-\frac{2m_k k_{f}}{3\pi^{2}}k^{2}\Pi_{0}) ]^{-1} \nonumber \\
& +2W_{2}^{(\alpha)} k_{f}^{2}\left(2k^{2}\Pi_{0}-\Pi_{2}\right) 
\end{eqnarray}
where $\Pi_{2}$ e $\Pi_{4}$ are generalized Lindhard  functions given in \cite{Gar.ea:92},
and we have introduced the adimensional variables $k=\frac{q}{2k_{f}}$ and $\nu=\frac{m_k E}{\hbar^{2}qk_{f}}$,
expressed in terms of the Fermi momentum $k_F$ and of the effective mass $m_k$.
The coefficients $W_{i}^{(\alpha)}$ 
can be expressed in terms of the usual coefficients defining the Skyrme interactions.
In neutron matter one finds
\begin{eqnarray}
W_1^{0}&=& t_0 (1-x_0) + \frac{1}{4}t_1 q^2 (1-x_1) -\frac{3}{4} t_2 q^2 (1+x_2)  \nonumber \\
& &+ \frac{t_3}{6} \rho^{\gamma} (1-x_3) \frac{(\gamma+1)(\gamma+2)}{2}
\nonumber \\
W_1^1&=& -t_0 (1-x_0) - \frac{1}{4}t_1 q^2 (1-x_1) -\frac{1}{4} t_2 q^2 (1+x_2)    \nonumber \\
& & -\frac{t_3}{6} \rho^{\gamma} (1-x_3)   \nonumber \\ 
W_2^0&=&  \frac{1}{4} t_1 (1-x_1) + \frac{3}{4} t_2 (1+x_2)   \nonumber \\
W_2^1&=&  -\frac{1}{4} t_1 (1-x_1) + \frac{1}{4} t_2 (1+x_2)   
\end{eqnarray}

The unperturbed response is obtained substituting 
the Lindhard function $\Pi_{0}$ in place of $\Pi(q,E)$ .
 
We shall compare these analytic results  with the numerical calculations performed 
putting 498 neutrons in a WS cell of radius $R_{WS} = 42.2$ fm without protons. In the latter case  
the response function  reads
\begin{equation}\label{NS comp: eq 20}
  S^{(\alpha)}_{WS}(q,E)=\sum_{n=0}^{+\infty}{\mid < n\mid e^{i\overrightarrow{q}\cdot\overrightarrow{r}} 
\Theta^{(\alpha)} \mid0 > \mid}^2 L(E,E_{n}),	
\end{equation}
where $\ket{0}$ is the ground-state, $\ket{n}$ are the (mean-field or RPA) excited states of the system,
and 
the computed discrete response has been convoluted with a Lorentzian function $L(E,E_n)$.

Exploiting the spherical symmetry of the system, one can introduce the multipole
decomposition of the external operator 
\begin{equation}\label{NS comp: eq 22}
  e^{i\vec{q}\cdot\vec{r}}=
         4\pi\sum_{L=0}^{+\infty}\sum_{M=-L}^{+L}i^{L}j_{L}(qr)Y_{LM}^{*}(\Omega_{q})Y_{LM}(\Omega_{r}),
\end{equation}
where $j_{L}(qr)$ is a spherical Bessel function, leading to an analogous  decomposition
of the response in the WS cell:
\begin{equation}\label{NS comp: eq 24}
  S^{(\alpha)}_{WS}(q,E)=\sum_{L=0}^{+\infty} S^{(\alpha)}_{WS,L}(q,E). \nonumber 
\end{equation}
The RPA excited states 
are given by 
\begin{equation}\label{NS comp: eq 28}
\begin{array}{l}
  \ket{n}\equiv \ket{\nu_{JM}} = \\
\\
\sum_{ph} \left(X_{ph}^{\nu}(J) \ket{p(h)^{-1}} +Y_{ph}^{\nu}(J) (-)^{J+M} \ket{(p)^{-1}h}\right),
\end{array}
\end{equation}
where the $X$ and $Y$ denote the forward and backward amplitudes 
calculated in RPA, 
and one obtains for density modes $(S=0,J=L)$
\begin{equation}\label{NS comp: eq 28.2}
 \begin{array}{l}
 S^{(0)}_{WS,L} (q,E) = 4\pi (2L+1) \sum_{\nu=0}   \left|  \sum_{j_{p}j_{h}} \left( X_{ph}^{\nu}(L) + \right. \right. \\
\\
\quad \left. \left.  (-)^{L} Y_{ph}^{\nu}(L) \right )\bra{j_{p}}|i^{L} Y_{L} |\ket{j_{h}} \bra{R_p} | j_{L}(qr) | \ket{R_h} \right|^{2} L(E,E_{\nu}),
 \end{array}
\end{equation}

\begin{figure}[!h]
\centering
\includegraphics*[width=0.38\textwidth]{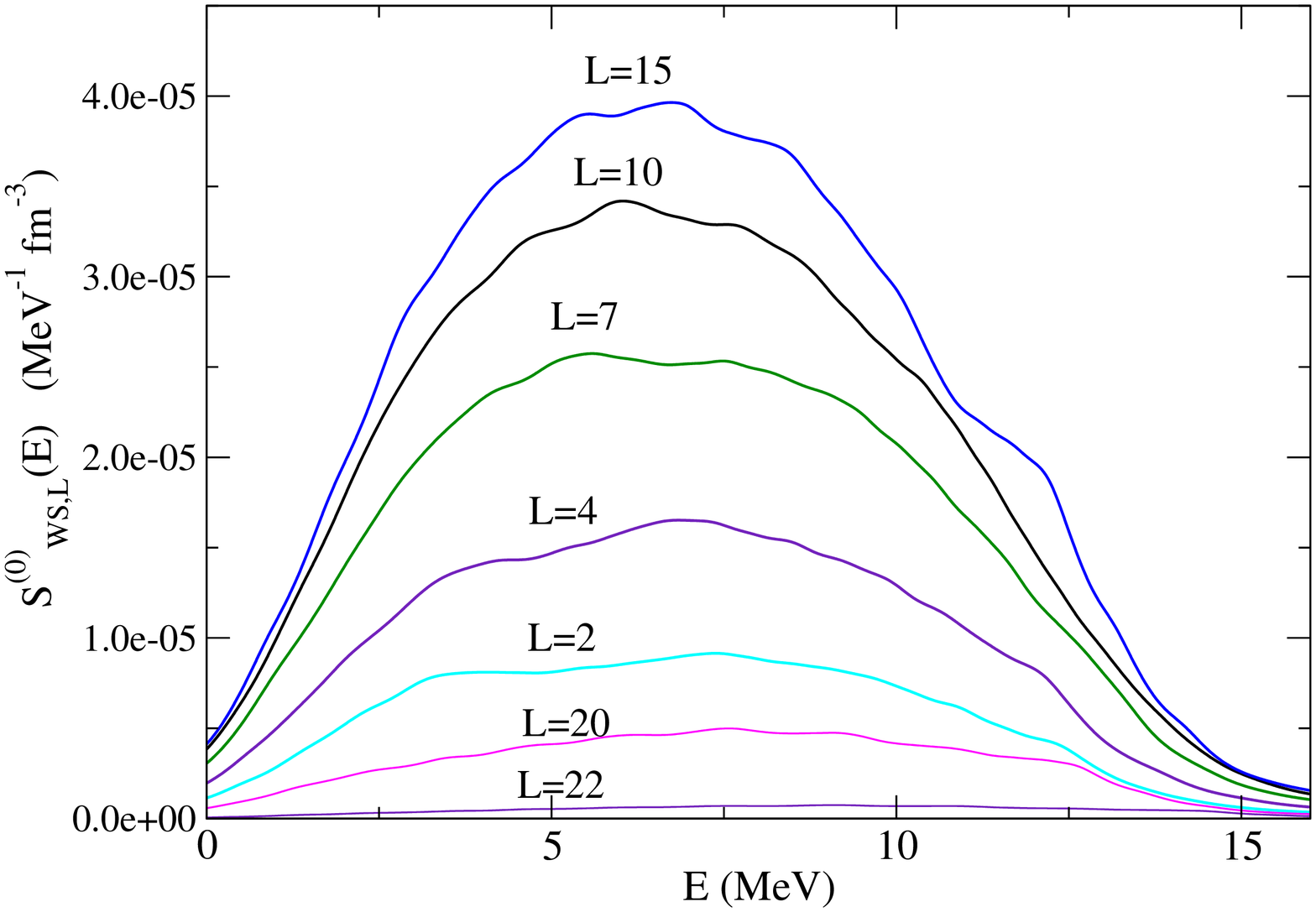}
\caption{Response function per unit volume as a function of the energy in the $S=0$ channel, computed in a
WS cell of 42.2 fm radius containing 498 neutrons.
The  contributions from  several multipolarities are shown.
The linear momentum transferred by the external field is $q$ = 0.51 fm$^{-1}$.}
\label{NS comp: fig 20}
\end{figure}

The contributions to the $S=0$ response in the cell associated
with several multipolarities 
are shown in   Fig. \ref{NS comp: fig 20}, for the momentum transfer $q = 0.51$ fm$^{-1}$. 
It is seen that the contributions from the different multipolarities increase with $L$ reaching
the maximum for $L=15$, and then rapidly decrease, becoming negligible for  
$L$ larger than about 20. This is in keeping with  
the simple estimate of the maximum 
angular momentum which can be transferred by the external field,  given by
$L_{max}\sim q R_{WS} \sim 21.$

\begin{figure}[!h]
\centering
\includegraphics*[width=0.38\textwidth]{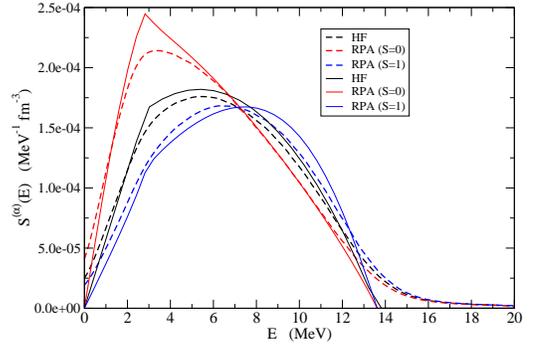}
\caption{Response functions  per unit volume as a function of the energy
in the $S=0$ and $S=1$ channel, computed in a WS
cell of  42.2 fm radius cell containing 498 neutrons
(dashed lines) and in uniform neutron matter (solid lines)
for the linear momentum $q$ = 0.51 fm$^{-1}$ transferred from the external field. 
Both systems correspond to a Fermi momentum  $k_{F}=$ 0.39 fm$^{-1}$. }
\label{NS comp: fig 10}
\end{figure}

The total response $S^{WS}(q,E)$ to the external fields (\ref{external}) 
is compared in Fig. \ref{NS comp: fig 10} with the analytic result,
in uniform neutron matter at the same density ($\rho \sim 0.01  \rho_0$)  and 
for the same  value of $q$ =0.51 fm$^{-1}$,
both in the unperturbed and in the RPA case.
We find  reasonable agreement
with the calculations in the WS cell. The strength functions computed in the WS cell
display a tail, which is not present in the uniform system, where the excitation energy of the
system has a maximum value, imposed by energy and momentum conservation and equal to
\begin{equation}\label{NS comp: eq 30}
  \hbar\omega=\frac{\hbar^{2}}{2m_k}(q^{2}+2qk_{f}),
\end{equation}
which, in the present case, corresponds to about 13.5 MeV, taking into account that
the neutron effective mass is very close to the bare mass ($m_k = 0.99\;  m)$
at the present density.
The analytic Energy Weighted Sum Rule (EWSR)  associated with the unperturbed response function $S(q,E)$
is given by  
\begin{equation}\label{NS comp: eq 32}
\left(\frac{EWSR_{HF}}{V}\right)_{unif}=\rho\frac{\hbar^{2} q^{2}}{2m_k},
\end{equation}
while the RPA response function in the $S=0$ channel is given by 
 \begin{equation}\label{NS comp: eq 34}
\left(\frac{EWSR^{S=0}_{RPA}}{V}\right)_{unif}=\rho\frac{\hbar^{2}q^{2}}{2m}.
\end{equation}

\begin{figure}[!h]
\includegraphics[width=0.38\textwidth,angle=-90]{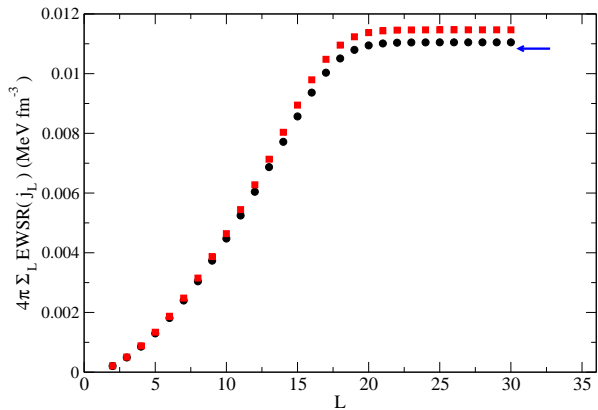} 
\caption{EWSR per unit volume calculated in the $S=0$ channel for as WS cell of 42.2 fm radius containing 508 neutrons, 
which corresponds to a Fermi energy  $k_{F}=$ 0.39 fm$^{-1}$. 
The linear momentum transferred from the external field $q$ = 0.51 fm$^{-1}$.
The squares show the cumulative EWSR  
obtained from the RPA strength function as a function of the orbital angular momentum $L$,
while the dots are obtained using Eq. (\ref{NS comp: eq 34.4}). 
The arrow
indicates the analytic value for the uniform system 
(cf. Eq. (\ref{NS comp: eq 34})).
}
\label{NS comp: fig 40}
\end{figure}

The Energy Weighted Sum Rule (EWSR) in the WS cell is numerically
calculated as the sum
of the EWSR associated to different multipolarities $J$:
\begin{equation}\label{NS comp: eq 34.2}
  EWSR\left(e^{i\vec{q}\cdot\vec{r}}\right) =4\pi \sum_{L}EWSR(j_{L}(qr)).
\end{equation}
The contribution calculated for each $L$ can be compared to the value 
obtained evaluating the classical EWSR on the HF ground state:
\begin{equation}\label{NS comp: eq 34.4}
\begin{array}{l}
EWSR_{classical}(j_{L}(qr))= \\
\\
\frac{\hbar^{2}}{2m} \frac{2\lambda+1}{4\pi} N <0|\left(\left(\frac{d }{d r}j_{L}(qr)\right)^{2}+L(L+1)\left(\frac{1}{r}j_{L}(qr)\right)^{2}\right)|0>,
\end{array}
\end{equation}
The cumulative EWSR is shown
in Fig. \ref{NS comp: fig 40}
for the RPA calculation. The total value
and turns out to be only a few per cent larger than the analytic value.

We conclude that   the numerically  computed response in the WS cell without protons
is quite close to the response calculated analytically in uniform neutron matter at the same density.
This gives us confidence in our numerical approach, which is  
applied in the main text  to  the case of a WS cell with  a nucleus in its center.

\section{Appendix B}


In ref. \cite{Khan,Grasso}, the strength associated with the operator $r^L$
was calculated in the inner crust in RPA and in QRPA (see also \cite{Gori}). Usually the operator
$r^L$ is used as a convenient approximation to the operator $e^{i\vec q \cdot \vec r}$ 
in the limit
of long wavelengths. In the inner crust  one is dealing with boxes with of the
order of 20 fm and with Fermi energies of the order of several MeV,   so that 
the produce $k_F R $ is not small and the choice of this operator does not seem
to be well motivated. 

\begin{figure*}
\begin{center}
\includegraphics[width=0.34\textwidth]{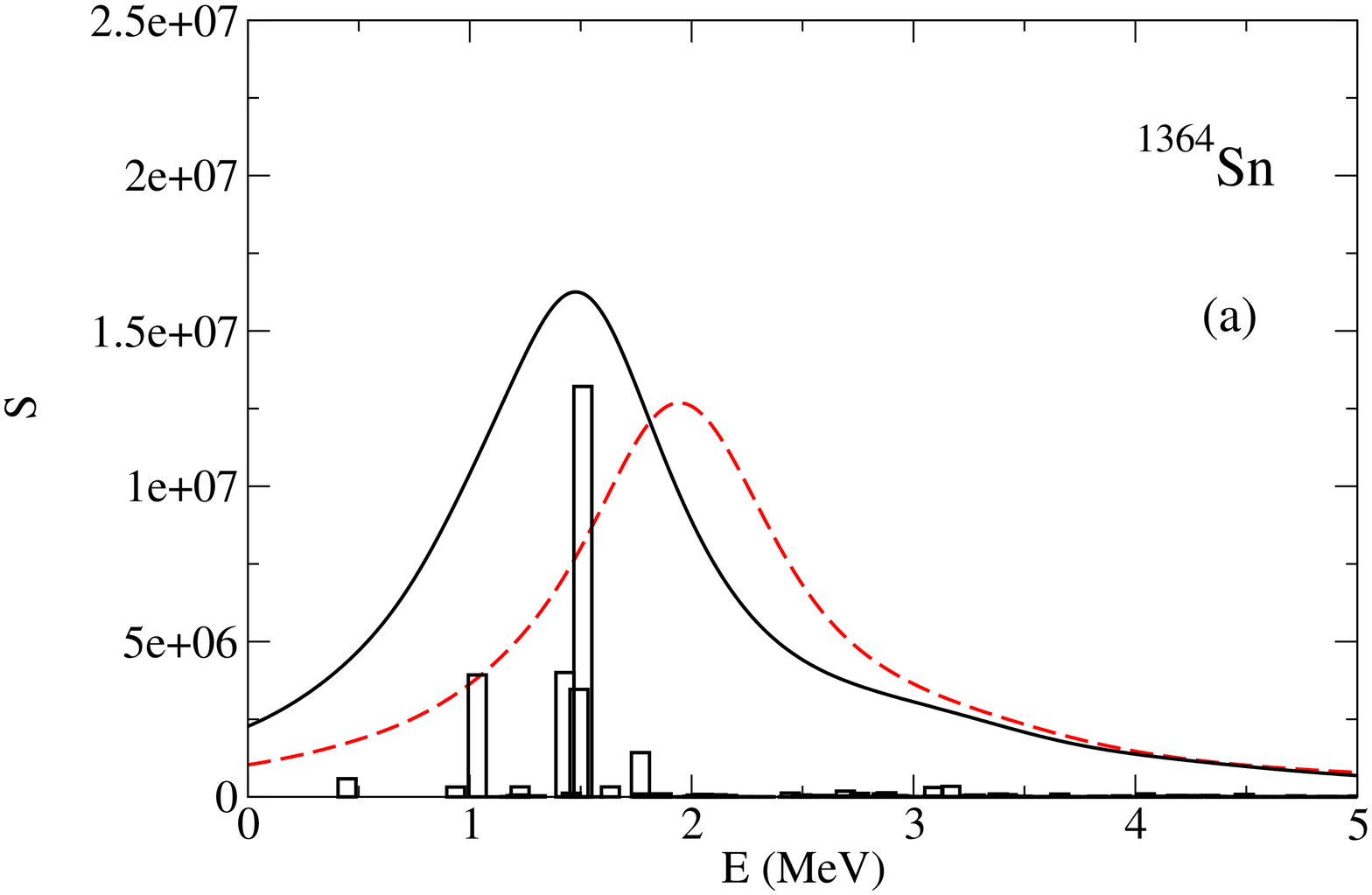}
\includegraphics[width=0.34\textwidth]{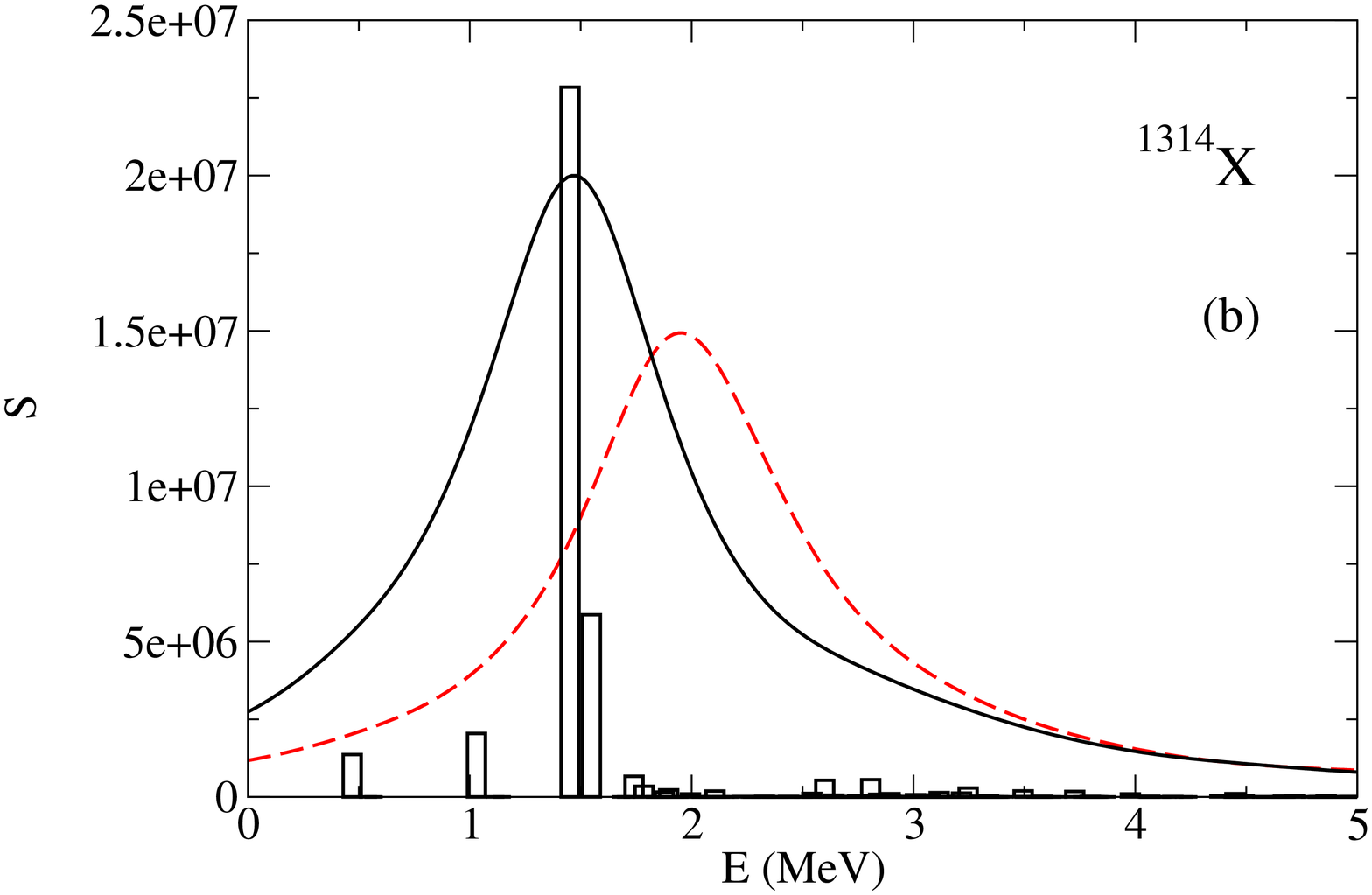}\\
\vspace{1cm}
\includegraphics[width=0.34\textwidth]{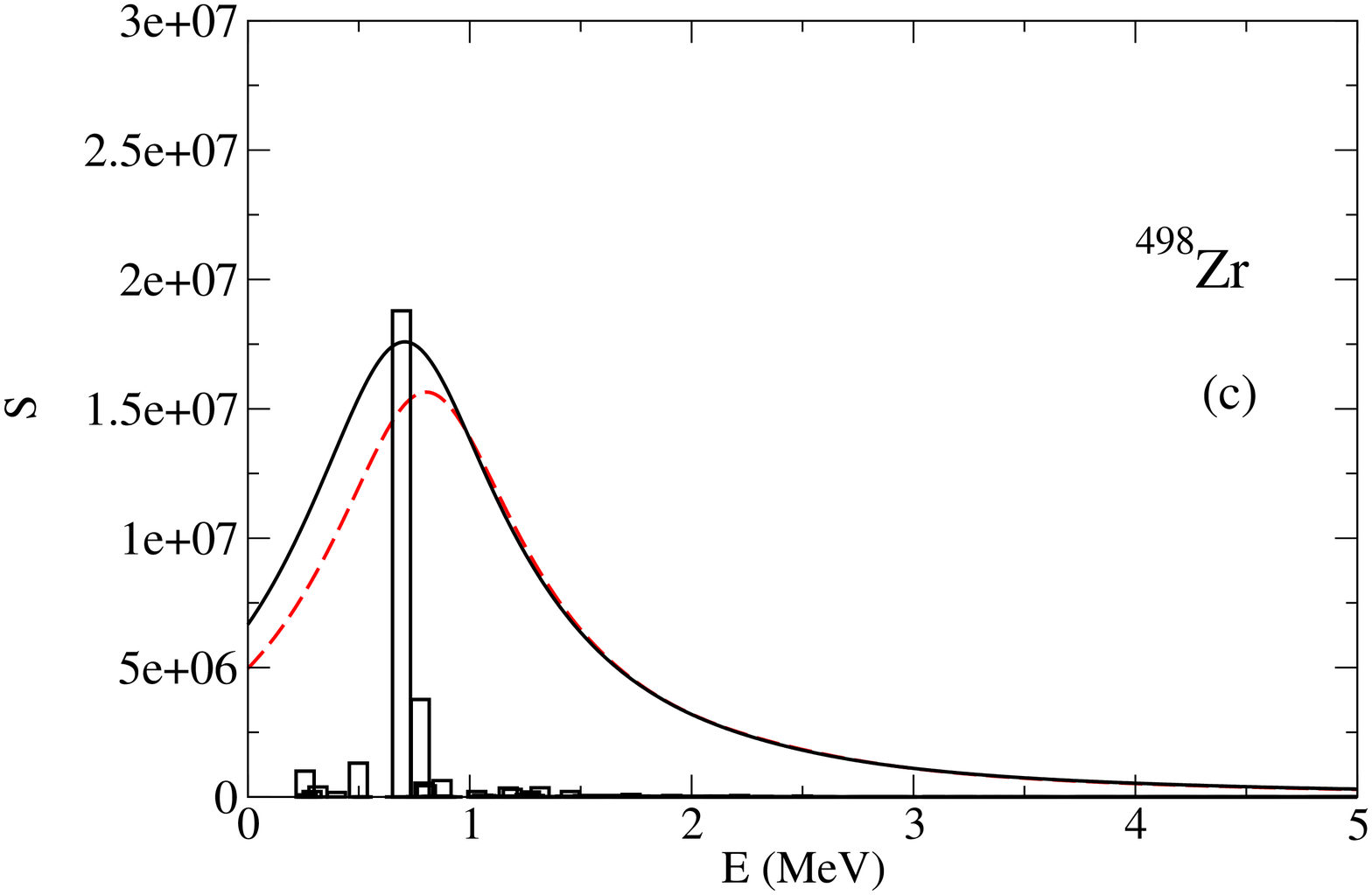}
\includegraphics[width=0.34\textwidth]{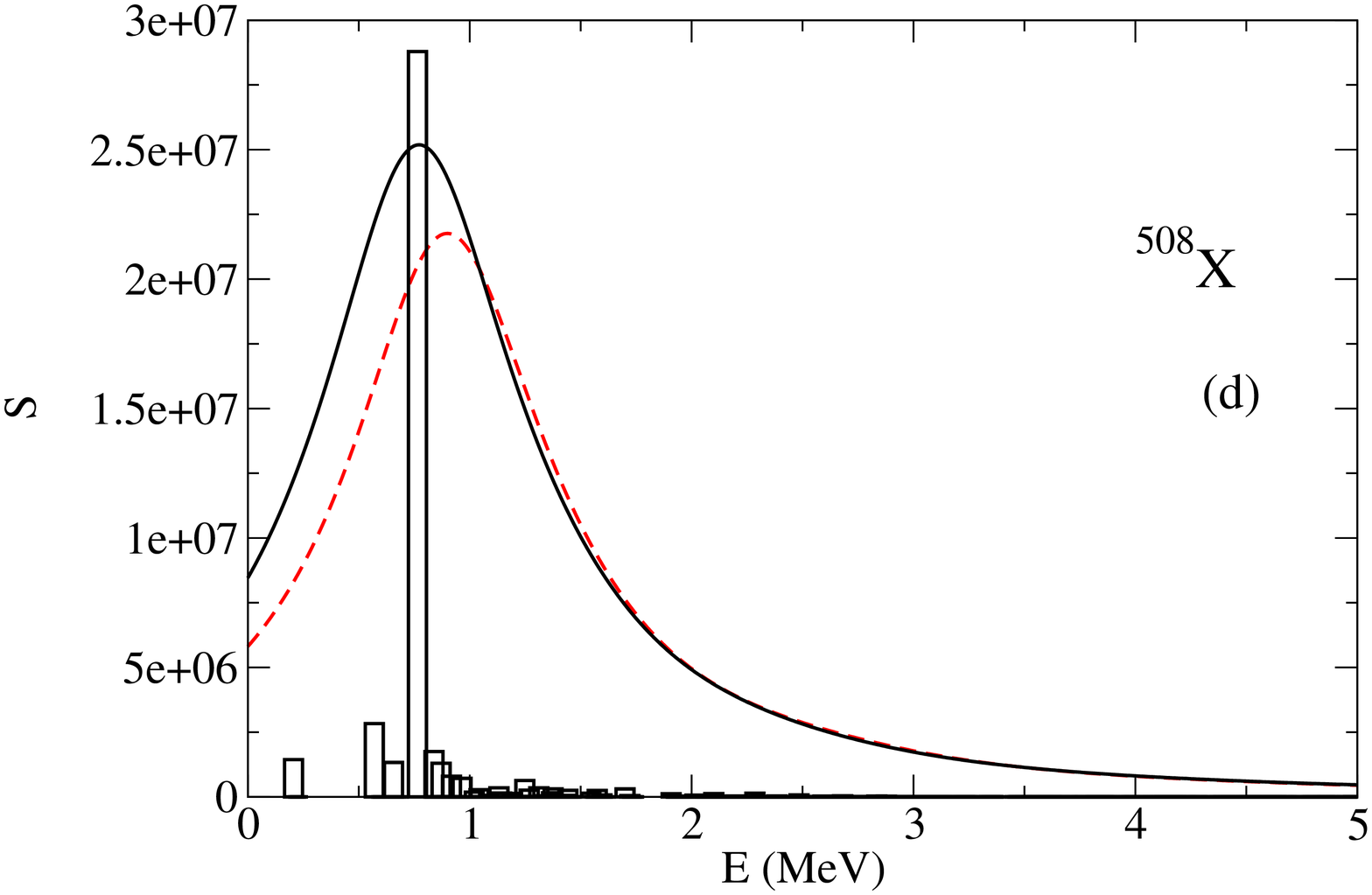}
\caption{(top) HF and RPA quadrupole strength functions calculated in the cells
$^{1364}$Sn (a) and $^{1314}$X (b), in units of MeV$^{-1}$ fm$^4$. 
(bottom) The same but for the cell $^{498}$Zr (c) and $^{500}$X (d).
The transition strengths
of the discrete states calculated in RPA are also shown in histogram form
(in units of fm$^4$).
}
\label{strength_rl}
\end{center}
\end{figure*}

Nevertheless, we have calculated the strength function associated with 
$r^2 Y_{2M}$  for the cells   $^{1364}$Sn  and $^{498}$Zr 
in order to compare with the results obtained in \cite{Grasso}.
The strength functions are shown in Fig. \ref{strength_rl}, where   
we also show calculations for two cells without protons
and with a similar number of nucleons, indicated as  $^{1314}$X
and $^{508}$X. 
In the present case, a cut off $E_{cut} =30 $ MeV is sufficient to exhaust the EWSR.

\begin{figure*}
\begin{center}
\includegraphics[width=0.36\textwidth]{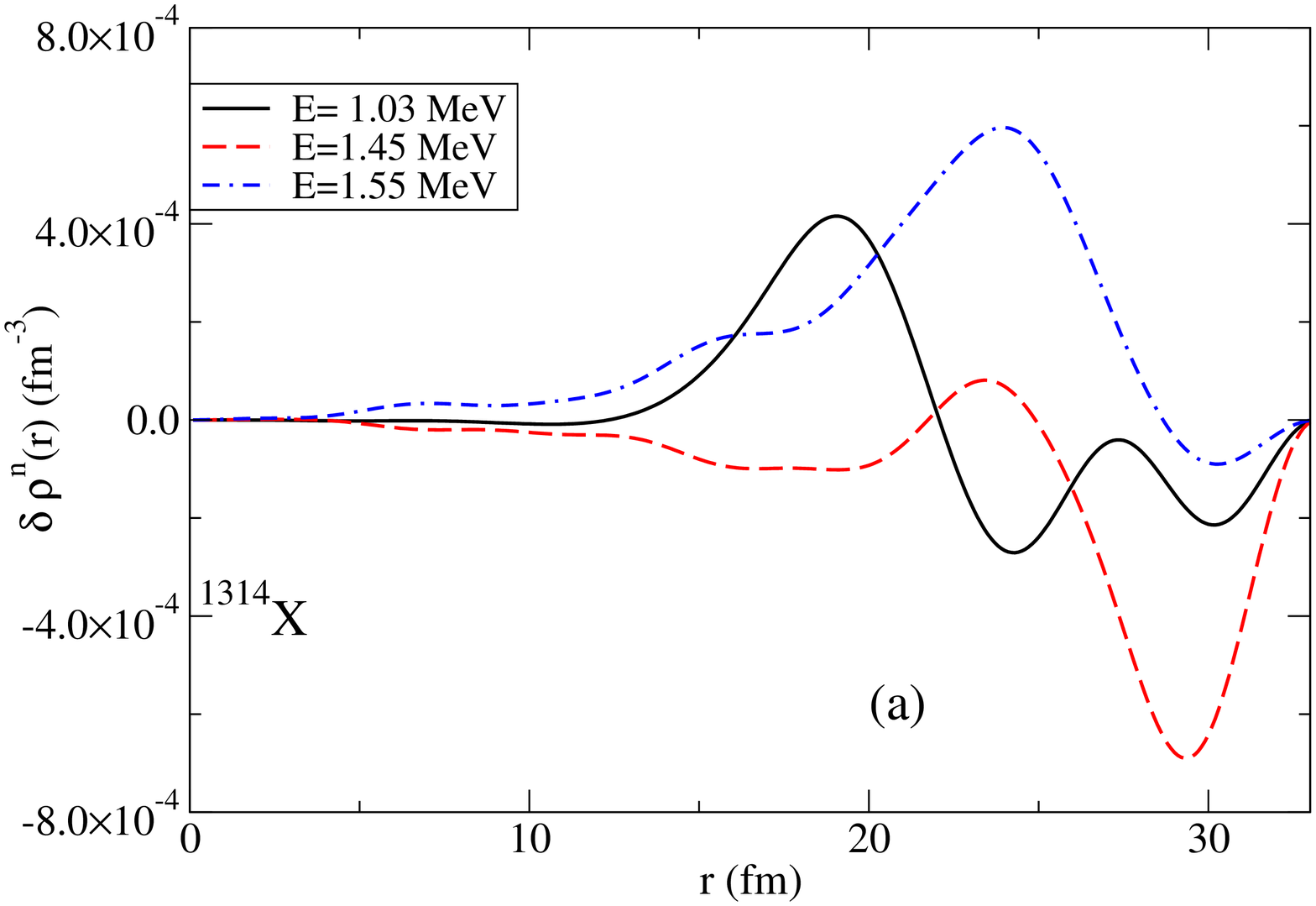}
\hspace{1cm}
\includegraphics[width=0.36\textwidth]{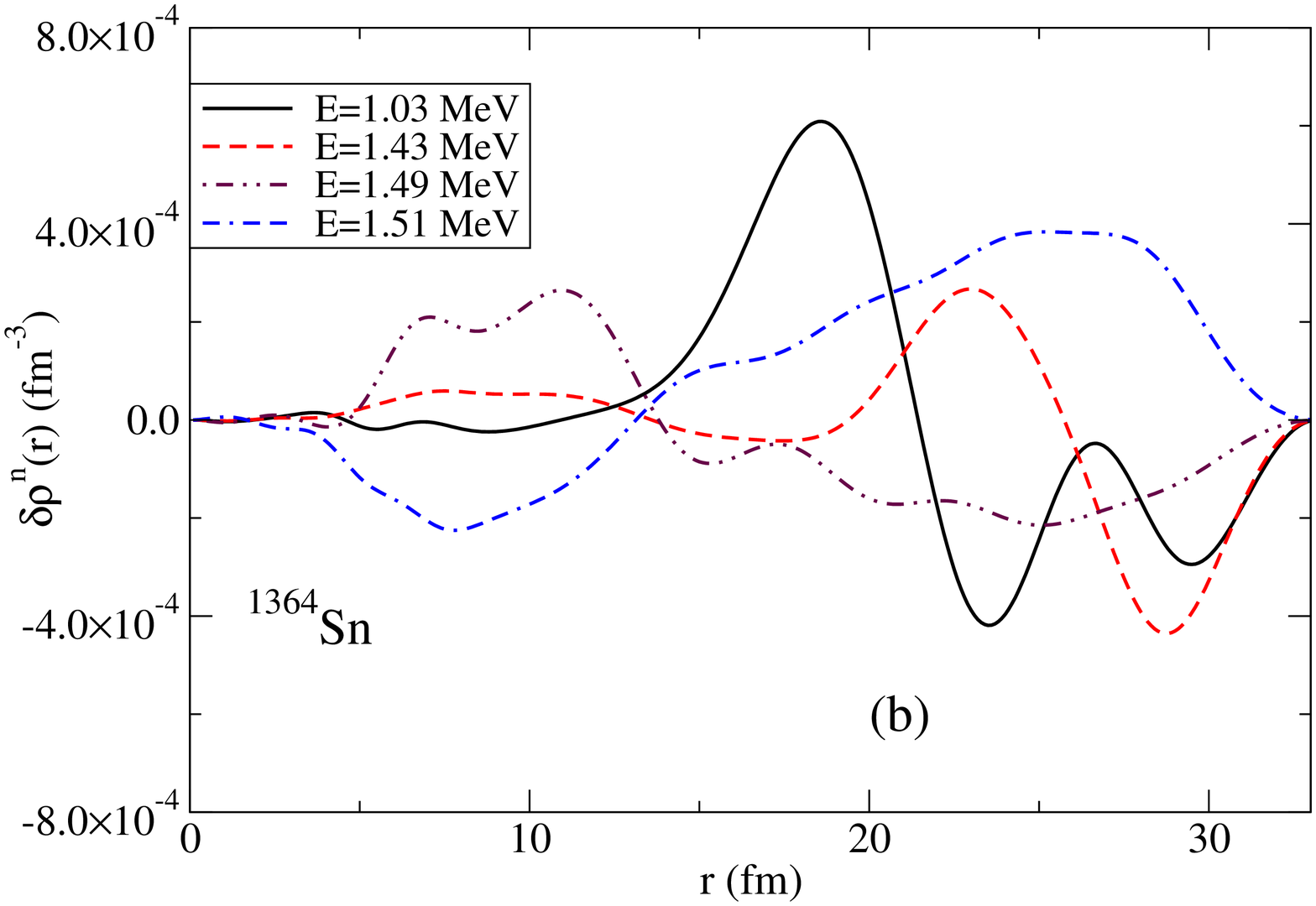}
\caption{Neutron transition densities associated with the strongest phonons 
calculated in $^{1314}$X (a) and in $^{1364}$Sn (b). The energy of the phonons is 
indicated. }
\label{trans_rl}
\end{center}
\end{figure*}

Our results coincide with those calculated by the Orsay group  
for the same cells \cite{Grasso-priv}, 
and the strength function show the features discussed in refs. \cite{Khan} and 
\cite{Grasso}, displaying a strong bump at low energy
('named supergiant resonance' in refs. \cite{Khan,Grasso}). The presence of the 
nucleus does not change the energy dependence of the strength, 
except for the fine details of 
the fragmentation of the discrete peaks produced by Landau damping.
The main effect of the 
nucleus is the decrease of the absolute value of the EWSR: the EWSR for   $^{1314}$X is about 20\%   
smaller than for $^{1364}$Sn, due to the fact that in the latter
about 150 neutrons are bound in the 
nucleus, and couple  to the $r^2$ operator much less efficiently than the 
unbound neutrons.
The neutron transition densities associated with the vibrations bulding the main peak
are shown in Fig. \ref{trans_rl}, for the $^{1314}$X and for the $^{1364}$Sn cell.
In both cases they show large values in the middle of the cell. The nucleus   
in the $^{1364}$Sn case only induces a modest increase of the transition densities
for $r \approx$ 10 fm for two of the phonons.   These results are  obviously 
in keeping  with the fact that the $r^2$ operator  weights heavily the region
far from the nucleus, at variance from  the $dU/dr Y_{LM}$  operator considered in the main text.

\clearpage

\bibliographystyle{apsrev}

\end{document}